\documentclass[aps,pra,preprint,epsfigure,superscriptaddress, amsmath,amssymb]{revtex4-1}
\usepackage[colorlinks=true,linkcolor=blue,urlcolor=blue,citecolor=blue,pdfusetitle]{hyperref}
\usepackage[utf8]{inputenc}
\usepackage[english]{babel}
\usepackage{amsmath}
\usepackage[caption = false]{subfig}
\usepackage{graphicx,epstopdf}
\usepackage{blindtext}
\usepackage[table,xcdraw]{xcolor}
\usepackage{lipsum}
\usepackage{amsfonts}
\usepackage{bbm}
\usepackage{amssymb}
\usepackage{enumerate}
\usepackage{color}
\usepackage{latexsym}
\usepackage{physics}
\usepackage{times,txfonts}
\usepackage{threeparttable}
\usepackage{graphicx}
\usepackage{subfig}
\usepackage{ulem}

\newtheorem{theorem}{Theorem}

\usepackage[ragged]{sidecap}

\newcommand{\Ocal}{\mathcal{O}}

\newcommand{\Scal}{\mathcal{S}}

\newcommand{\Rmath}{\mathbbm{R}}

\newcommand{\SubFig}[2]{\ref{#1}{\color{blue}#2}}

\definecolor{bluePoli}{cmyk}{0.4,0.1,0,0.4}
\definecolor{blueGreen}{RGB}{0, 102, 102}
\definecolor{brickred}{rgb}{0.8, 0.25, 0.33}
\definecolor{darkred}{RGB}{204, 0, 0}
\definecolor{darkgreen}{RGB}{0, 102, 50}
\definecolor{darkblue}{RGB}{0, 76, 153}
\definecolor{myorange}{RGB}{204, 0, 200}

\newcommand{\revisionRefA}[1]{{\color{black}#1}}
\newcommand{\revisionRefB}[1]{{\color{black}#1}}
\newcommand{\revisionRefC}[1]{{\color{black}#1}}

\newcommand{\iqa}{International Quantum Academy, Futian District, Shenzhen, Guangdong 518055, China}
\newcommand{\siqse}{Shenzhen Institute for Quantum Science and Engineering, Southern University of Science and Technology, Shenzhen, Guangdong 518055, China}
\newcommand{\physustech}{{Department of Physics, Southern University of Science and Technology, Shenzhen, Guangdong 518055, China}}
\newcommand{\gdpkl}{Guangdong Provincial Key Laboratory of Quantum Science and Engineering, Southern University of Science and Technology, Shenzhen, Guangdong 518055, China}
\newcommand{\szkl}{Shenzhen Branch, Hefei National Laboratory, Shenzhen 518048, China}
\newcommand{\DFAU}{{Department of Physics and Astronomy,~Aarhus University,~Ny munkegade 120, 8000 Aarhus C, Denmark}}

\newcommand{\UFSCar}{Departamento de F\'{i}sica, Universidade Federal de S\~ao Carlos, S\~ao Carlos 13565-905, São Paulo, Brazil}
\newcommand{\CSIC}{Instituto de Física Fundamental, Consejo Superior de Investigaciones Científicas, Calle Serrano 113b, 28006 Madrid, Spain}
\newcommand{\SYU}{State Key Laboratory of Precision Spectroscopy, School of Physics and Material Science, East China Normal University, Shanghai 200062, China}

\newcommand{\Title}{Digital simulation \revisionRefC{of zero-temperature spontaneous} symmetry breaking in a \\ superconducting lattice processor}

\usepackage{orcidlink}

\begin{document}

\title{\Title}

\author{Chang-Kang Hu}
\affiliation{\iqa}\affiliation{\siqse}\affiliation{\gdpkl}

\author{Guixu Xie}
\affiliation{\iqa}\affiliation{\siqse}\affiliation{\gdpkl}

\author{Kasper Poulsen}
\affiliation{\DFAU}

\author{Yuxuan Zhou}
\affiliation{\iqa}\affiliation{\siqse}\affiliation{\gdpkl}

\author{Ji Chu	}
\affiliation{\iqa}\affiliation{\siqse}\affiliation{\gdpkl}

\author{Chilong Liu}
\affiliation{\iqa}\affiliation{\siqse}\affiliation{\gdpkl}

\author{Ruiyang Zhou}
\affiliation{\iqa}\affiliation{\siqse}\affiliation{\gdpkl}

\author{Haolan Yuan}
\affiliation{\iqa}\affiliation{\siqse}\affiliation{\gdpkl}

\author{Yuecheng Shen}
\affiliation{\SYU}

\author{Song Liu}
\email{lius3@sustech.edu.cn}
\affiliation{\iqa}\affiliation{\siqse}\affiliation{\gdpkl}\affiliation{\szkl}

\author{Nikolaj T. Zinner}
\affiliation{\DFAU}

\author{Dian Tan}
\email{tand@sustech.edu.cn}
\affiliation{\iqa}\affiliation{\siqse}\affiliation{\gdpkl}

\author{Alan C. Santos~\orcidlink{0000-0002-6989-7958}}
\email{ac\_santos@iff.csic.es}
\affiliation{\CSIC}
\affiliation{\UFSCar}

\author{Dapeng Yu}
\email{yudp@sustech.edu.cn}
\affiliation{\iqa}\affiliation{\siqse}\affiliation{\gdpkl}\affiliation{\szkl}\affiliation{\physustech}


\begin{abstract}
	\textbf{Quantum simulators are ideal platforms to investigate quantum phenomena that are inaccessible through conventional means, such as the limited resources of classical computers to address large quantum systems or due to constraints imposed by fundamental laws of nature. Here, \revisionRefC{through a digitized adiabatic evolution,} we report an experimental \revisionRefB{simulation} of \revisionRefC{antiferromagnetic (AFM) and ferromagnetic (FM) phase formation induced by spontaneous} symmetry breaking (\revisionRefC{SSB}) in a three-generation Cayley tree-like superconducting lattice. \revisionRefC{We develop a digital quantum annealing algorithm to mimic the system dynamics, and observe the emergence \revisionRefC{of signatures of SSB-induced phase transition} through a connected correlation function. We demonstrate that the signature of \revisionRefC{phase transition from classical AFM to quantum FM} happens in systems undergoing zero-temperature adiabatic evolution with only nearest-neighbor interacting systems, the shortest range of interaction possible.} By harnessing properties of the bipartite Rényi entropy as an entanglement witness, we observe the formation of \revisionRefC{entangled quantum FM and AFM phases}. Our results open perspectives for new advances in condensed matter physics and digitized quantum annealing.} 
\end{abstract}

\maketitle


\subsection*{\large Introduction}

Symmetry in physical systems has led to a number of scientific and technological advances related to conservation laws of nature encapsulated by Noether's theorem~\cite{Noether1918}. On the other hand, symmetry breaking is a key mechanism in condensed matter physics and the standard model~\cite{Higgs:64,Guralnik:64,Sigrist:91,Anderson:Book}, and second-generation quantum devices in spintronics~\cite{Kasper:PRL22,Kannan:23,Diez:23}. \revisionRefC{In particular, at finite-temperature systems, spontaneous symmetry breaking (\revisionRefC{SSB}) is of great interest for quantum phase transitions phenomena in low-dimensional systems \revisionRefB{(one and two dimensions)}~\cite{Sadler:06,Sachdev:Book,Lu:17,Trenkwalder:16}, as it may lead to formation of genuine long-range order when the physical system contains sufficiently long-ranged interactions~\cite{MerminWagnerTheorem,Bruno:01}}. Recently, two independent experimental realizations in one-dimensional trapped ion chains~\cite{Feng:23} and two-dimensional Rydberg atoms arrays~\cite{Chen:23} have successfully observed continuous symmetry breaking and the formation of long-range order. \revisionRefB{Experimental realizations in trapped ions have been possible because such a phenomena in low-dimensional quantum systems can be achieved for power-law interactions as $V(r) \sim r^{-\alpha}$, with $\alpha < 3$~\cite{Maghrebi:17}.} \revisionRefC{Although the  Mermin-Wagner~\cite{MerminWagnerTheorem,Bruno:01}, which forbids formation of correlated antiferromagnetic (AFM) and ferromagnetic (FM) states, does not apply to zero-temperature systems,} it is believed that \revisionRefC{SSB} is also forbidden for one-dimensional systems~\cite{Maghrebi:17}. Therefore, this subject is less explored than its finite-temperature counterpart~\cite{MerminWagnerTheorem,Bruno:01,Feng:23,Chen:23,Maghrebi:17,Block:22}. 

In this scenario, gate-based digital quantum simulators can efficiently observe zero-temperature quantum phenomena or processes, since such a cooling regime for quantum analog computers is not allowed as a consequence of the third law of thermodynamics: the unattainability principle~\cite{nernst1906ueber,Masanes:17}. As a promising platform for both digitized and analog tasks, superconducting integrated circuits are universal platforms to \textit{mimic} adiabatically driven quantum processes at zero temperature through quantum annealing~\cite{King:21,King:23}, or digitized adiabatic evolutions (DAE)~\cite{Barends:16}. The DAE method aims to create a model of computation that takes advantage of two different models: adiabatic quantum computation~\cite{Farhi:01} and quantum circuit model of computation~\cite{Benioff:80,Benioff:82,Deutsch:85}. Inspired by the application potential of DAE, the goals of this work are twofold: i) establish a general and sufficient condition for high-fidelity digitized adiabatic quantum computation, and ii) report the first experimental simulation of zero-temperature \revisionRefC{SSB} in a \revisionRefB{two-dimensional (2D)} Cayley tree spin chain \textit{without} long-range interactions. The signature of \revisionRefC{SSB} is observed through two-point correlation functions and the second-order Rényi entropy, revealing the correlation profile over the system and the emergence of genuine entanglement in the system, respectively.

\subsection*{\large The digital quantum annealing}

\begin{figure}[h!]
	\includegraphics[width=\columnwidth]{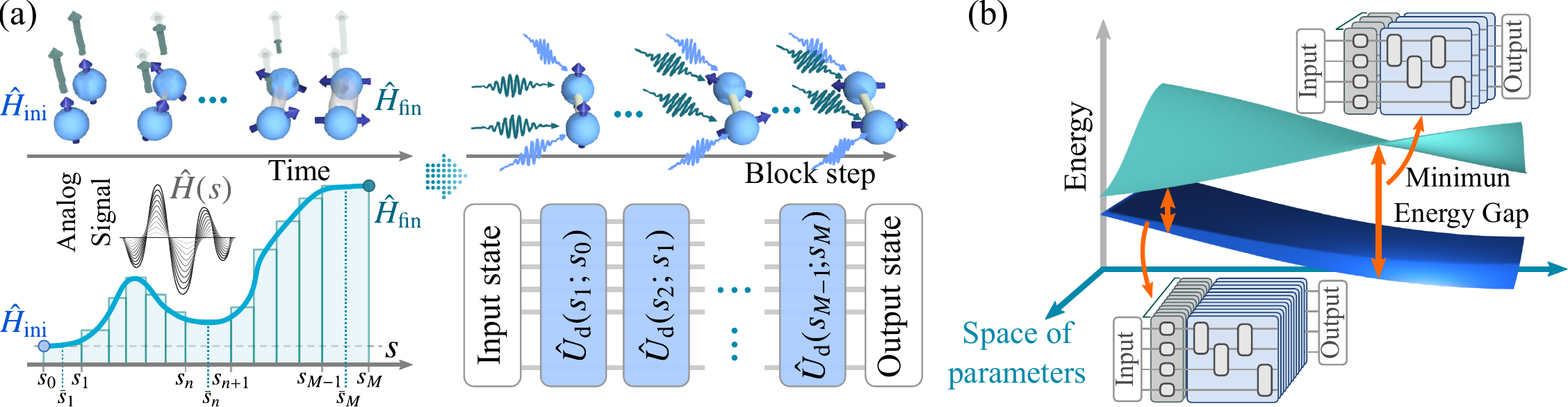}
	\caption{\textbf{Digitization and adiabatic energy gap.} (a) The procedure to digitize an adiabatic evolution is done through a Riemann-like discretization of the time interval $s \in [0,1]$, where each step in time corresponds to the digital block. The time-continuous adiabatic algorithm implemented through time-dependent fields can be efficiently decomposed in a sequence of pulses through a circuit version of the evolution. After $M$ blocks the output state is expected to be prepared with good fidelity without any computation complexity due to the search for the optimal parameters of the circuit.	
		(b) The only optimization required to reduce the circuit length is done through the suitable choice of the parameters of the Hamiltonian. The \textit{a priori} knowledge of the parameters of the Hamiltonian,   which leads to a large energy gap, will enhance the digitized algorithm.}
	\label{Fig:SchemeFig}
\end{figure}

Since the zero-temperature \revisionRefC{SSB} transition is simulated by keeping the maximum purity of the system, this can be done through closed system adiabatic evolution~\cite{Kato:50}. For this reason, as sketched in Fig.~\SubFig{Fig:SchemeFig}{a}, we aim to investigate the digital adiabatic evolution driven by the time-dependent Hamiltonian of the generic form $\hat{H}(t) = f(t) \hat{H}_{\text{ini}} + g(t) \hat{H}_{\mathrm{fin}}$. The functions $f,g$ satisfy $f(0)=g(\tau)\neq 0$ and $f(\tau)=g(0) = 0$, with $\tau$ the total evolution time. In this way, we are able to describe how the experimental simulation of \revisionRefC{SSB} for the lattice is engineered through digitization of adiabatic evolution. To this end, we first introduce the \revisionRefC{ \textit{Suzuki-Trotter adiabatic digitization}} that governs the performance of adiabatic quantum optimization tasks in digital quantum processors. Given the Schrödinger equation, $i \hbar \ket*{\dot{\psi}(t)} = \hat{H}(t)\ket{\psi(t)}$, to digitize the dynamics we consider the Riemann-like discretization of the evolution, as sketched in Fig.~\SubFig{Fig:SchemeFig}{a}. Let us define now the normalized time $s \in [0,1]$ as $s = t/\tau$. In this approach, the evolution for the $n$-th step of the digitizing procedure is governed by the evolution operator $\hat{U}_{\mathrm{d}}(s_{n+1};s_{n})$, during a time interval $\delta s_{n} = s_{n+1} - s_{n}$. $\hat{U}_{\mathrm{d}}(s_{n+1};s_{n})$ is obtained from the Hamiltonian parameters at the instant of time $\bar{s}_{n} = (s_{n+1} + s_{n})/2$ and assuming a time-independent evolution for a time duration of $\delta s_{n}$ (first approximation of the method). Following this strategy, we aim to find a quantum circuit able to properly mimic the evolution, where each operator $\hat{U}_{\mathrm{d}}(s_{n+1};s_{n})$ is given by the first-order Trotter decomposition (second approximation). In this way, \revisionRefC{as shown in Methods section, we can properly digitize the evolution and make a direct relation between the adiabatic energy gap and the total number of blocks $M$ required for the digitized annealing procedure through the digitized decomposition of the evolution operator as
\begin{equation}
	\hat{U}_{\mathrm{d}}(s_{n+1};s_{n}) \approx
	e^{- \frac{i\tau \delta s_{n}}{\hbar}g(\bar{s}_{n}) \hat{H}_{\mathrm{fin}} }e^{- \frac{i\tau \delta s_{n}}{\hbar}f(\bar{s}_{n}) \hat{H}_{\mathrm{ini}} } . \label{Eq:Approximation}
\end{equation}

In this way, as depicted in Fig.~\SubFig{Fig:SchemeFig}{b}, the performance of the digitized adiabatic evolution may be optimized by finding the best functions $f(s)$ and $g(s)$ to maximize the minimum adiabatic gap. This can be done by a suitable choice of the Hamiltonian interpolation functions as determined by quantum adiabatic brachistochrone trajectories~\cite{Rezakhani:09,Santos:21a}, which will be considered in this work.}

\revisionRefB{It is worth to mention that, even though this method is not a variational algorithm and classical optimization is not required, the complexity of the problem may be as costly as simulating $\hat{H}_{\mathrm{fin}}$. In general, such a complexity will increase with the number of qubits and the presence of many-body terms in $\hat{H}_{\mathrm{fin}}$, and in this sense the digital annealing is as complex as any other simulation or variational algorithm. However, when $\hat{H}_{\mathrm{fin}}$ is efficiently implemented in a real quantum processor, the complexity of the digital algorithm, i.e. the number of blocks $M$, does not necessarily increase. In this case, we can take advantage of digital annealers even for larger systems. To exemplify this, in the Supplementary Material~\cite{SM} we show the numerical simulation of a digitized SSB in a 15 qubits using a number of blocks \textit{smaller} than that one required for SSB in a 7 qubits system.}

\subsection*{\large The three-generation Cayley tree-like device}

\begin{figure}[h!]
	\includegraphics[width=0.85\columnwidth]{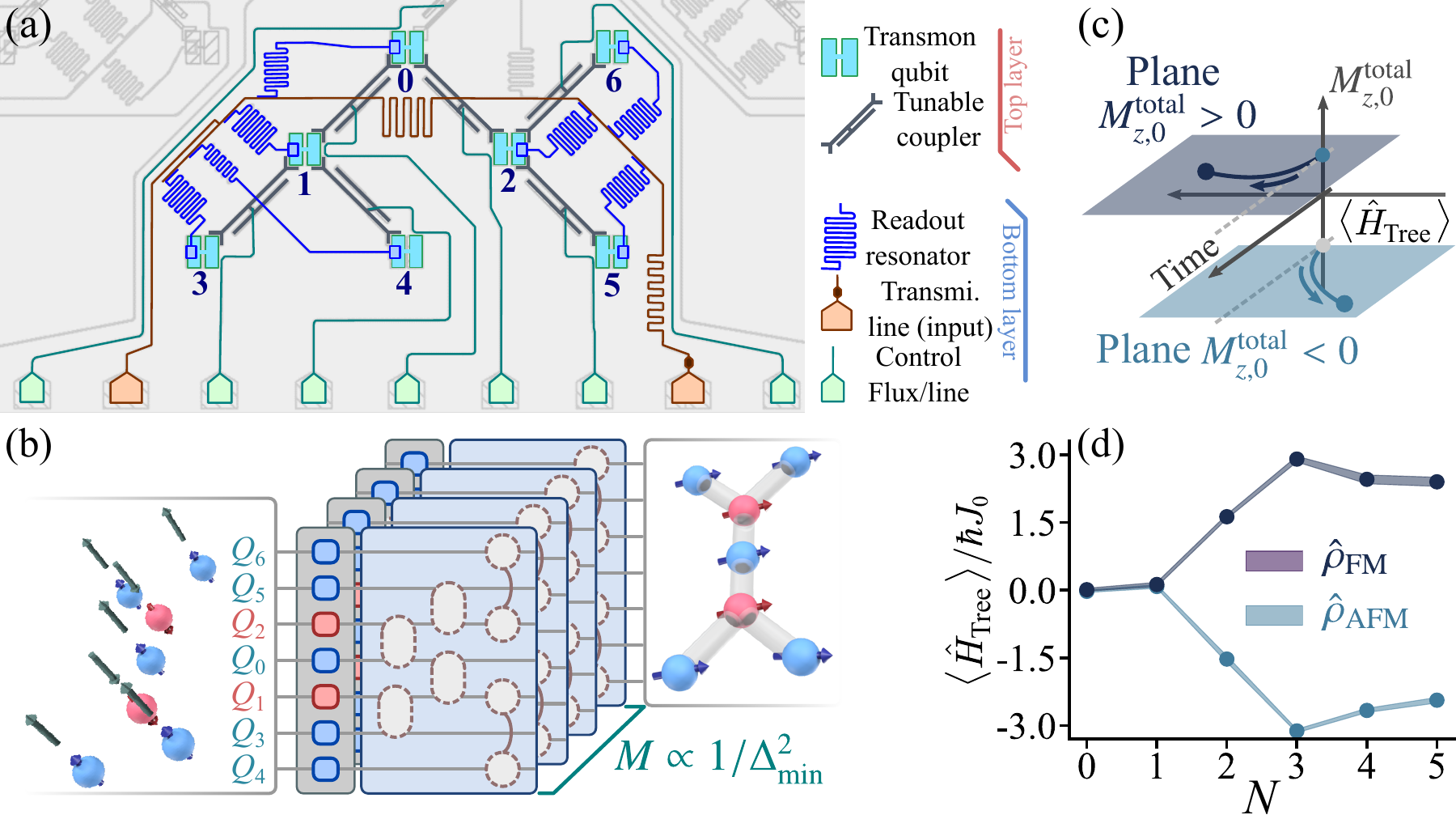}
	\caption{\textbf{Device setup, circuit and phase-dependent energy split.} (a) The 7-qubit Tree sub-lattice used in our experiment. Our chip is a 3D chip, in which the transmission line, readout resonators, and flux and control lines are placed on a layer different from the couplers and transmon qubits. 
		(b) The circuit considered in our experiment is composed of a first layer of single-qubit gates, emulating local fields, followed by interaction blocks to mimic the spin-spin interaction during the evolution. As stated by Theorem~\ref{Theorem}, given the minimum energy gap $\Delta_\mathrm{min}$, high fidelity digitization is achieved when $M \propto 1/\Delta_\mathrm{min}^2$. For the dynamics under consideration, we used $M = 5$ for all the simulations.
		In (c) we sketch the influence of the symmetry on the dynamics, where the transition into FM and AFM quantum phases occurs in different magnetization planes. 
		The panel (d) is the experimental data for the energy splitting along the digitized evolution of the system from the initial to the final state. Time is encoded in the number of blocks of the digitized circuit ($N=0$ and $N=5$ refers to $t=0$ and $t=\tau$, respectively), for $J\tau=5$.}
	\label{Fig:SchemeFig2}
\end{figure}

The physical system of interest in the experimental setup consists of a three-generation Cayley tree-like superconducting \revisionRefB{2D-lattice} shown in Fig.~\SubFig{Fig:SchemeFig2}{a}. Our tree-like processor employs a flip-chip packaging process in which the device consists of a 3D chip with two layers of superconducting elements, separated by $9$ micrometers. In this setup, the qubits and couplers are placed on the top layer of the chip, while the complex of transmission and control lines and readout resonators are placed on the bottom layer. This pristine environment enhances the quality of both qubits and couplers with respect to undesired systematic errors. These seven qubits exhibited an average single-qubit gate fidelity of $99.92\%$, and for two-qubit gate we reached 98.68\%. More information on the device parameters and calibration is presented in Supplementary Material~\cite{SM}.

\revisionRefB{In our experiment the quantum processor is not a zero-temperature system, as it is impossible by fundamental laws of nature, and the temperature of the quantum processing unity is around 10~mK~\cite{SM}. However, the target evolution we will implement using an experimental digital quantum circuit corresponds to a unitary evolution of the system driven by an ideal adiabatic Hamiltonian at zero temperature. As we shall see, even under presence of noise, gate errors and thermal fluctuations, we can properly observe the emergence of SSB from our experimental digital simulation because it in fact capture the zero-temperature aspect of the target evolution.}

The system is driven through a \revisionRefC{SSB}-induced transition from an initial state given by the classical Néel state. \revisionRefB{In our system, the Néel state  is defined in such a way that the spin state of any pair of adjacent spins are opposite to each other, that is, they are initially at state $\ket{\uparrow \downarrow}$ or $\ket{\downarrow\uparrow}$ (as shown in Fig.~\SubFig{Fig:SchemeFig2}{b}). For our topology, such Néel states can be obtained from the generation-staggered field Hamiltonian for the tree lattice given by~\footnote{It is worth to mention that our initial Hamiltonian is not unique, as any other Hamiltonian that commutes with $H_\mathrm{Neel}$ share the same set of eigenstates $H_\mathrm{Neel}$. However, if we assume only the set of all non-interacting Hamiltonians, then the Hamiltonian $\hat{H}_{\text{Néel}}$ considered is \textit{unique}.}}
\begin{equation}
	\hat{H}_{\text{Néel}} = \hbar\sum\nolimits_{l=0}^{L-1} \sum\nolimits_{n_{l}=1}^{N_{l}}  (-1)^l \omega_0 \hat{\sigma}_{n_{l}}^{z} , \label{Eq:HamiltBNT}
\end{equation}
where $L$ denotes the total number of generations of the tree (for $7$ spins, one has $L=3$) and $N_{l}$ the total number of spins in the $l$-th generation, given in terms of $l$ as $N_{l} = 2^l$.  The above Néel Hamiltonian has two important energy states, namely, the ground and excited classical Néel states given, respectively, by
\begin{equation}
	\ket*{\psi_{\mathrm{ground}}^{\text{Néel}}} = \ket{\downarrow}_{1^{\mathrm{st}}}\ket{\uparrow}_{2^{\mathrm{nd}}}\ket{ \downarrow}_{3^{\mathrm{rd}}}, \quad \ket*{\psi_{\mathrm{excited}}^{\text{Néel}}} = \ket{ \uparrow}_{1^{\mathrm{st}}}\ket{ \downarrow}_{2^{\mathrm{nd}}}\ket{ \uparrow}_{3^{\mathrm{rd}}} , \label{Eq:IniState}
\end{equation}
where $\ket{ \uparrow}_{n^{\mathrm{th}}}$ denotes that all spins in the $n$-th layer are in the spin-up state, and similarly for $\ket{ \downarrow}_{n^{\mathrm{th}}}$. The highest energy eigenstate of the Hamiltonian $\hat{H}_{\text{Néel}}$ is denoted the excited classical Neel state. In this way, a given spin site with positively-oriented local magnetic field only interacts with a negatively-oriented one, when the interaction Hamiltonian for our system device reads
\begin{equation}
	\hat{H}_{\mathrm{Tree}} =	\hbar \sum\nolimits_{\langle n, k \rangle} J_0 \left(\hat{\sigma}_{n}^{+}\hat{\sigma}_{k}^{-}+\hat{\sigma}_{n}^{-}\hat{\sigma}_{k}^{+}\right) , \label{Eq:H_bethe}
\end{equation}
where $\sum_{\langle n, k \rangle}$ is a sum over all connections of the tree-like lattice. Therefore, the adiabatic Hamiltonian for the \revisionRefC{SSB} in our device is $\hat{H}(s) = f(s)\hat{H}_{\text{Néel}} + g(s)\hat{H}_{\mathrm{Tree}}$. \revisionRefB{As we shall see soon, the preparation of the system in the classical Néel states in Eq.~\eqref{Eq:IniState} are relevant to the final target state of the evolution because they allow us to have different quantum phases encoded in the correlated eigenstates of the final Hamiltonian $\hat{H}_{\mathrm{Tree}}$.}

All information about the circuit that digitizes $\hat{H}(s)$, such as connectivity and gate parameters, is obtained from the adiabatic digitization at first order Suzuki-Trotter, Eq.~\eqref{Eq:Approximation}. In our particular case, the native circuit topology (Fig.~\SubFig{Fig:SchemeFig2}{a}) allows us to find an optimized gate sequence of the circuit to simulate each block~\cite{SM}. Our gate parameters are chosen to approximate the Quantum Adiabatic Brachistochrone interpolation functions~\cite{Rezakhani:09}
\begin{align}
	f_\mathrm{bra}(s) = 1 - \frac{1}{2}\left[ 1 - \tan \left( \frac{(1-2s)\pi}{4} \right) \right] , \quad g_\mathrm{bra}(s) = \frac{1}{2}\left[ 1 - \tan \left( \frac{(1-2s)\pi}{4} \right) \right] .
\end{align}

A preliminary simulation of the analog adiabatic dynamics indicates that a total evolution time given by $\tau J_0 = 5$ is enough to approximately achieve the adiabatic regime of this dynamics. After each block of the circuit, shown in Fig.~\SubFig{Fig:SchemeFig2}{b}, we measure physical quantities related to the outcome state of the system, focusing on the signature of \revisionRefC{SSB} transitions through two complementary quantities: energy and pairwise connected correlations.

The system has a symmetry in $\hat{M}_{z}$, since $[\hat{H}_{\text{Néel}},\hat{M}_{z}]=0$, therefore each adiabatic evolution from the ground/excited Néel states will take place over a different magnetization ($\langle\hat{M}_{z}\rangle$) plane (Fig.~\SubFig{Fig:SchemeFig2}{c}). The expectation value of the total magnetization, $\hat{M}_{z} = (\hbar/2) \sum_{n}\hat{\sigma}_{z}^{n}$, for the ground (exited) state is negative (positive). Therefore, as a first quantity able to highlight the transition from the classical AFM Néel state to quantum AFM/FM phases, we evaluate the energy of the system with respect to the Tree Hamiltonian $\langle \hat{H}_{\mathrm{Tree}} \rangle$. For this purpose, we use $\langle \hat{H}_{\mathrm{Tree}} \rangle/\hbar  J_0 = \sum\nolimits_{\langle n, k \rangle} \big( \langle\hat{\sigma}_{n}^{x}\hat{\sigma}_{k}^{x}\rangle + \langle\hat{\sigma}_{n}^{y}\hat{\sigma}_{k}^{y} \rangle \big)$, where the two-body $\langle\hat{\sigma}_{n}^{x/y}\hat{\sigma}_{k}^{x/y}\rangle$ terms are experimentally obtained by standard measurement protocols~\cite{Nielsen:Book}. As it can be seen \revisionRefB{from the experimental data in} Fig.~\SubFig{Fig:SchemeFig2}{d}, both the ground and excited Néel states have the same energy with respect to this reference Hamiltonian, namely $\langle\hat{H}_{\mathrm{Tree}}\rangle = 0$. However, by driving the system continuously from the initial Néel field Hamiltonian $\hat{H}_{\text{Néel}}$ to the tree-like one $\hat{H}_{\mathrm{Tree}}$, the instantaneous system energy, with respect to the $\hat{H}_{\mathrm{Tree}}$, presents an energy splitting due to the formation of either the FM or the AFM phase, depending on the initial Néel state considered (sketched in Fig.~\SubFig{Fig:SchemeFig2}{d}). When the system starts in the classical AFM Néel state with negative magnetization (ground state), it follows a trajectory on the negative magnetization plane while decreasing the system energy. Conversely, by preparing the system in the AFM Néel state with positive magnetization (excited state), we observe a spontaneous transition of the system to the (ordered) quantum FM state, which leads to an increase in its energy while its magnetization keeps constant during the process.

\revisionRefB{We take advantage of the Cayley tree-like lattice to observe the SSB phenomena using the minimum amount of gates as possible. It is possible because our lattice is two-dimensional, but the number of interactions in the system is significantly smaller than a square lattice, for example. In this way, we only need to simulate a small number of interactions, which allows us to efficiently observe SSB with a few blocks digitized evolution. In fact,} the topology of our superconducting device permits an efficient gate sequence of the circuit to simulate each block. The  simulation of spin-spin interaction is done through two-qubit CZ gates and single qubit rotations as depicted in Fig.~\SubFig{Fig:CSB}{a}, with two parameters to be determined $\phi^{0}_{z,n}$ and $\varphi_{J,n}$. For arbitrary functions $f,g\in \Rmath$, the parameters of the circuit, which implements the \revisionRefC{evolution}, are immediately obtained from $\phi^{l}_{z,n} = (-1)^{l} \omega_{0} \tau f(\bar{s_{n}}) \delta s_{n}$, and $\varphi_{J,n} = J_{0} \tau g(\bar{s_{n}}) \delta s_{n}$, where $\phi^{l}_{z,n}$ and $\varphi_{J,n}$ are dimensionless parameters associated to the initial local fields and interaction terms of the Hamiltonian, respectively. We simplify our circuit by defining a single parameter $\phi^{0}_{z,n}$ such that $\phi^{l}_{z,n} = (-1)^{l}\phi^{0}_{z,n}$.

\subsection*{\large \revisionRefC{Phase} transition signature and formation of entangled quantum phases}

We measure the two-point correlation function, which is defined as $C_{x}^{(i,j)} = \langle \hat{\sigma}^{x}_{i}\hat{\sigma}^{x}_{j} \rangle - \langle \hat{\sigma}^{x}_{i} \rangle \langle \hat{\sigma}^{x}_{j} \rangle$. \revisionRefB{Because the isotropic aspect of the two-qubit correlations in the XY-plane, without loss of generality we show the correlations only along the $x$-direction (see Supplementary Material for further discussions).} Fig.~\SubFig{Fig:CSB}{b} shows the behavior of the two-dimensional profile of the correlations ($C_{x}^{(i,j)}$), with respect to the two-spin sites $(i,j)$, as a function of the $n$-th block in the digitized circuit. So, we state one of our main results: \textit{\revisionRefC{while ferromagnetism and antiferromagnetism formation} are forbidden for short-range interacting systems at any finite temperature~\cite{MerminWagnerTheorem}, these phases can be accessed through zero-temperature evolution by exploring adiabatically driven dynamics of a nearest-neighbor interacting spin lattice. More than that, we also observe signature of phase transition from uncorrelated classical AFM states to a correlated quantum FM phase.} In fact, the result shown in Figs.~\SubFig{Fig:CSB}{b},~\SubFig{Fig:CSB}{c} and~\SubFig{Fig:CSB}{d} is clear evidence of a dynamical symmetry breaking in the system. On the one hand, when we initialize the system in the ground state we see the emergence of the AFM phase of the XY Hamiltonian. On the other hand, by starting the system in the excited state, we achieve a final state consistent with a quantum ordered FM phase. \revisionRefB{This behavior is consistent with a phase transition from the classical AFM Néel state~\cite{NeelState} to the correlated FM state of the XY Heisenberg Hamiltonian, which is the signature of spontaneous symmetry breaking induced during the evolution of the system.}

\revisionRefB{Because the final Hamiltonian admits the existence of symmetries, at the final of the evolution the system energy spectrum is expected to be doubly degenerate, at minimum. Therefore, we state now that the existence of accessible states other than the target states cannot be populated along the evolution, and then destroy the formation of the correlated phases of the matter. In fact, it can be properly addressed by exploiting the symmetry of the system with respect to the eigenstate parity, defined as the expected value of the operator $\hat{\Pi}_{z} = \prod_{l=0}^{L-1} \prod_{n_{l}=1}^{N_{l}} \hat{\sigma}_{n_{l}}^{z}$. As detailed in Supplementary Material, because $[\hat{\Pi}_{z},\hat{H}(s)]=0$, the conservation law for $\hat{\Pi}_{z}$ allows to efficiently address the final state because the well-defined parity of each initial state as considered in Eq.~\eqref{Eq:IniState}. Therefore, degenerated states with different parity cannot be mixed during the evolution.}

In order to quantify the impact of errors in the digitized circuit implementation, we \revisionRefB{compare} the ideal and experimental correlation matrices. \revisionRefB{We define the similarity between the theoretical and experimental data as $\Scal = 1 - \max_{j,i}|[C_{\mathrm{the}} - C_{\mathrm{exp}}]_{j,i}|/2$}, where $[X]_{j,i}$ is the element $(i,j)$ of a given matrix $X$. The matrix $C_{\mathrm{the}}$ is the theoretical prediction for the correlation matrix, with elements $C_{x}^{(i,j)}$, as provided by the numerical simulation of the digitized circuit, and $C_{\mathrm{exp}}$ is the corresponding correlation matrix computed with the experimental outcomes of the digitized circuit. As a result, the \revisionRefB{similarity $\Scal$} of the experimental realization with respect to the desired ideal result is shown in Fig.~\SubFig{Fig:CSB}{c}. \revisionRefB{It is worth to mention that the analysis of the similarity not related to state fidelity, as it only quantifies the quality of the experimental data with respect to the theoretical one. However, by combining the result shown in Fig.~\SubFig{Fig:CSB}{c}, with the energy estimate in Fig.~\SubFig{Fig:SchemeFig2}{b}, and the two-point correlation profile in Fig.~\SubFig{Fig:CSB}{b}, we can associate high values of $\Scal$ with SSB and AFM-FM phase transition in our dynamics.} 

\revisionRefB{Our results} lead to the conclusion that the experiment is mainly affected by the single and two-qubit gate errors. However, even under the influence of such undesired effects, it is worth highlighting that the digitization of our \revisionRefC{7-qubit scheme} provides a final \revisionRefB{accuracy} of around $80\%$. Additionally, \revisionRefB{as shown in Figs.~\SubFig{Fig:CSB}{b},~\SubFig{Fig:CSB}{c}, by properly choosing the final time of the evolution, and the digitized step (for example $n=3$ and $n=4$), the emergence of the \revisionRefC{phase transition} can be observed with enhanced sharpness. In others words, the formation of ordered quantum FM phase from the AFM classical state emerges early in the evolution ($s < 1.0$), so we could stop the digitized dynamics at $n=3$ or $n=4$ to reduce the accumulated errors. For example, the signature of \revisionRefC{SSB, as well as formation of correlated quantum FM and AFM phases,} can be efficiently} captured by the digital circuit after the second block with \revisionRefB{similarity} around $90\%$ \revisionRefB{(in case $n=2$). The good performance of the circuit decomposition around the middle of the evolution in Fig.~\ref{Fig:CSB}, also observed in Fig.~\SubFig{Fig:SchemeFig2}{d}), can be justified by the small variations in the quantum adiabatic brachistochrone interpolation function used in the experimental circuit, as detailed in SM~\cite{SM}}. This demonstrates the resilience of our digitized approach in simulating the relevant phenomenon under consideration.

We also investigate spin-spin correlation behavior as a function of the ``distance" between the spins (Fig.~\SubFig{Fig:CSB}{d}). To measure this quantity, we use the distance between two spins $j$ and $i$ given by the Manhattan distance $r_{ij} = |s_{j} - s_{i}|$, where the separation between two neighbor spins is $d$ and the reference spin is the first spin of the 2nd generation, namely, spin 3. The graphical view of the Manhattan distance is presented in Fig.~\SubFig{Fig:CSB}{d}, where our reference spin at the origin is highlighted. The main conclusion is that the behavior of the range of correlations in our Cayley tree device differs from the long-range model observed for linear and 2D lattices~\cite{Feng:23,Chen:23}, where the FM phase exhibits a correlation length bigger than the AFM one \revisionRefB{due to the nature of AFM and FM spin states and symmetry breaks~\cite{Peter:12,Maghrebi:17}}.

To highlight the quantumness of the FM and AFM phases observed in Fig.~\ref{Fig:CSB}, we also analyze the increased quantum correlation in the system. To this end, we observe the formation of entanglement entropy as witnessed by the second-order Rényi entropy, given by $S_{\text{Rényi}}(\hat{\rho}) = -\log_{2}\big[\mathrm{Tr}(\hat{\rho}^2)\big]$, for bi-partitions of the tree lattice encoded in our device. \revisionRefC{The Rényi entropy reveals aspects of inseparability for both pure and mixed quantum states~\cite{Horodecki:96}, therefore} it is used here as a witness of entanglement formation between two subsystems, say $A$ and $B$, of a given system $AB$. More precisely, by denoting the output density matrix $\hat{\rho}_{\mathrm{FM}/\mathrm{AFM}}$ for the FM/AFM phase, and the reduced density matrix $\hat{\rho}^{A}_{\mathrm{FM}/\mathrm{AFM}}$ of the subsystem A, $S_{\text{Rényi}}(\hat{\rho}_{\mathrm{FM}/\mathrm{AFM}}) < S_{\text{Rényi}}(\hat{\rho}^{A}_{\mathrm{FM}/\mathrm{AFM}})$ implies entanglement between the partition A and the rest of the system~\cite{Horodecki:96,Horodecki:96PLA,Horodecki:09}. \revisionRefC{In case of pure states we have $S_{\text{Rényi}}(\hat{\rho}_{\mathrm{FM}/\mathrm{AFM}}) = 0$, and therefore any quantity $S_{\text{Rényi}}(\hat{\rho}^{A}_{\mathrm{FM}/\mathrm{AFM}})>0$ implies entanglement.}

\begin{figure}[t]
	\includegraphics[width=\columnwidth]{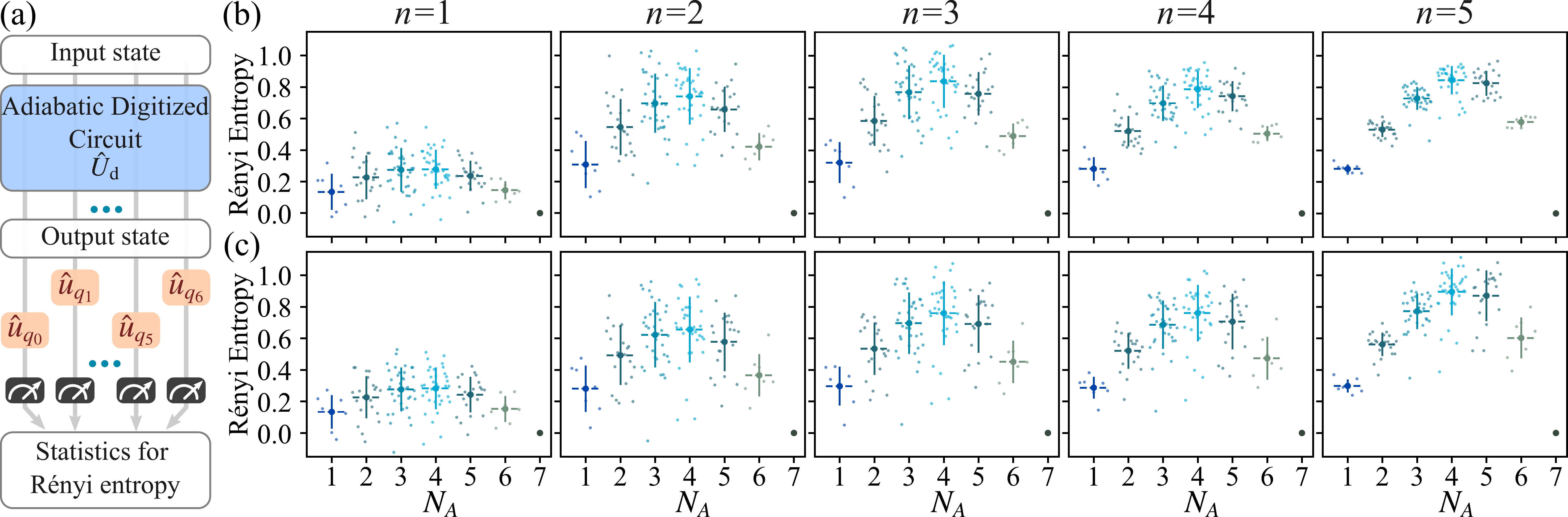}
	\caption{\textbf{Sketch for measurement of Rényi entropy of the system:} (a) The sequence of steps to measure the Rényi entropy is presented. After the adiabatic digitized circuit is implemented through the unitary digital operator $\hat{U}_{\mathrm{d}}$, we apply random single-qubit gates to the output state and perform the joint measurement of the whole system on the computational basis. (b, c) Bipartite quantum correlations (Rényi entropy) of different choices of the subsystem A for each block of the digitized protocol (from $n=0$ to $n=M=5$), in cases where the system is initialized in the Néel ground state (b) and the Néel exited state (c). Each set of points corresponds to the Rényi entropy for a choice of the $A$ subsystem obtained through the random measurements, where average values and their corresponding standard deviations are shown (horizontal and vertical bars).}
	\label{Fig:Renyi}
\end{figure}

The experimental evaluation of the Rényi entropy is done through randomized measurements~\cite{vanEnk:12,Elben:18,Tiff:19}, carried out in our work as follows. First, we characterize all possible combinations of subsystems with $N_{A}$ qubits in $A$ and $7-N_{A}$ qubits in B. After the experiment circuits, we apply a product of single-qubit unitaries to all seven of our qubits, denoted as $\hat{U} = \hat{u}_{q_{0}} \otimes \ldots \otimes \hat{u}_{q_{6}}$. Each unitary $\hat{u}_{q_{i}}$ is independently drawn from the Circular Unitary Ensemble (CUE)~\cite{Mezzadri:07}. Subsequently, we measure the qubits in the $\sigma_{z}$-basis (computational basis). We perform multiple joint measurements with 10000 shots on the whole system for each instance of $U$ to gather statistical data, and we repeat this entire process for 100 different randomly selected instances of $U$. Using the data obtained through the above process and employing the statistical analysis methods provided in the literature~\cite{Tiff:19}, we can obtain the second-order Rényi entropy $S_{\text{Rényi}}(\hat{\rho}^{A})$ of any subsystem A with $N_{A}$ spins (qubits), as depicted in Fig.~\SubFig{Fig:Renyi}.

The Rényi entropy after correcting the noise effects are seen in Fig.~\SubFig{Fig:Renyi}, where each ``cloud" of points in the graph denotes the Rényi entropy for each size $N_{A}$ of the subsystem A. For each value of $N_{A}$, we have $c(N_{A}) = 7!/N_{A}!(7-N_{A})!$ points in the cloud due to the number of combinations for the bipartite decomposition $A$-$B$. 
The average value and the standard deviation for each cloud are highlighted. 
This analysis exposes that the correlations spread almost identically over the system for both the AFM and FM phases from the top and bottom panels of Fig.~\ref{Fig:Renyi}, respectively, showing that the correlation range of both phases obeys the same decay profile. 
\revisionRefC{Even if the system is not pure state at the end of the experiment, but after correction of the noise effects, the Rényi entropy sill can be used as a witness of entanglement with the condition $S_{\text{Rényi}}(\hat{\rho}^{A}_{\mathrm{FM}/\mathrm{AFM}}) > S_{\text{Rényi}}(\hat{\rho}_{\mathrm{FM}/\mathrm{AFM}})$.}





\subsection*{\large Conclusion}

In this work, we have done theoretical and experimental investigations on zero-temperature SSB in a tree-like spin-spin interacting system with only nearest-neighbor interaction. While fundamental theorems forbid such a system to undergo a phase transition from the classical AFM Néel state to the quantum FM state induced by SSB at finite temperature, the results presented in the main text and Supplementary Material~\cite{SM} suggest such a phenomenon is possible through zero-temperature adiabatic evolution. The quantumness of the FM phase is witnessed through two-point correlation functions and the formation of entropy of entanglement as given by the second-order Rényi entropy. Since the applicability of our framework goes beyond the particular phenomenon studied here, we expect to observe SSB occurring for other geometries and structures, like regular 2D lattice, zigzag lattice, and among others. For quantum simulation and computation, the sufficient condition for high-fidelity DAE is useful to achieve high-fidelity for digital adiabatic-inspired algorithms. As an alternative to QAOA~\cite{Farhi:14,Farhi:22}, DAE constitutes a promising strategy to build quantum circuits for the optimization process, but without classical optimization required by hybrid models of optimization. Therefore, it establishes a route to perform optimization tasks in quantum processors that cannot be used as \textit{quantum annealers}. 

Our results instigate further prospects on the simulation of physical systems and processes in digitized adiabatic quantum simulators. \revisionRefB{The challenges of the scalability of digitized annealing are mainly related to the accumulating errors due to limited gate fidelity. However, by employing efficient error mitigation techniques~\cite{Kandala:19,Kim:23,Kim:23b} we expect to be able to extrapolate the application of digitized quantum annealing to larger quantum processors.}



\subsection*{\large Online content.} 

Discussions on methods, additional references, Nature Research reporting summaries, source data, extended data and supplementary information are available at [\textit{\color{blue}the link to be defined}].

\bibliography{mybib-URL.bib}
\bibliographystyle{PRA-WithTitle.bst}

\subsection*{\large Methods}

\subsection*{Digitized adiabatic theorem} \label{ApSec:DigitizingHad}

We discuss here how the optimization of DAQC can be done through strategies to find the optimal adiabatic functions $f(t)$ and $g(t)$ responsible for driving the Hamiltonian between the initial ($\hat{H}_{\mathrm{ini}}$) and problem Hamiltonians ($\hat{H}_{\mathrm{fin}}$) according to the equation
\begin{equation}
	\hat{H}(s) = f(s) \hat{H}_{\mathrm{ini}} + g(s) \hat{H}_{\mathrm{fin}} .
\end{equation}

In this case we can write
\begin{equation}
	U_{\mathrm{dig}}^{\mathrm{std}}(s_{n+1};s_{n}) = e^{- \frac{i\tau \delta s_{n}}{\hbar}\left(f(\bar{s}_{n}) \hat{H}_{\mathrm{ini}} + g(\bar{s}_{n}) \hat{H}_{\mathrm{fin}}\right) }.
\end{equation}

By assuming that $\delta s_{n}$ is small enough to get a good approximation of the above equation, here we need to make sure that $U_{\mathrm{dig}}^{\mathrm{std}}(s_{n+1};s_{n})$ can be decomposed in simple quantum gates. For example, it would be desirable to decompose the above unitary into two independent unitary associated to $\hat{H}_{\mathrm{ini}}$ and $\hat{H}_{\mathrm{fin}}$, we mean 
\begin{equation}
e^{- \frac{i\tau \delta s_{n}}{\hbar}\left(f(\bar{s}_{n}) \hat{H}_{\mathrm{ini}} + g(\bar{s}_{n}) \hat{H}_{\mathrm{fin}}\right)}\approx
e^{- \frac{i\tau \delta s_{n}}{\hbar}g(\bar{s}_{n}) \hat{H}_{\mathrm{fin}} }e^{- \frac{i\tau \delta s_{n}}{\hbar}f(\bar{s}_{n}) \hat{H}_{\mathrm{ini}} } , \label{ApEq:Approximation}
\end{equation}
with good approximation. However, because $[\hat{H}_{\mathrm{ini}},\hat{H}_{\mathrm{fin}}]\neq 0$, such a decomposition is not possible in general. To this end, here we develop a general condition over $\delta s_{n}$ in order to get such a decomposition. We use the Baker-Campbell-Hausdorff to write
\begin{equation}
e^{- \frac{i\tau \delta s_{n}}{\hbar}g(\bar{s}_{n}) \hat{H}_{\mathrm{fin}} }e^{- \frac{i\tau \delta s_{n}}{\hbar}f(\bar{s}_{n}) \hat{H}_{\mathrm{ini}} } =
 e^{- \frac{i\tau \delta s_{n}}{\hbar}\left(f(\bar{s}_{n}) \hat{H}_{\mathrm{ini}} + g(\bar{s}_{n}) \hat{H}_{\mathrm{fin}}\right) 
+ \frac{(i\tau \delta s_{n})^2}{2 \hbar^2}g(\bar{s}_{n})f(\bar{s}_{n}) \left[\hat{H}_{\mathrm{fin}} ,\hat{H}_{\mathrm{ini}} \right] + \Ocal\left((\tau \delta s_{n})^3\right)} .
\end{equation}

From this equation, it is intuitive to say that if
\begin{equation}
\frac{(\tau \delta s_{n})^2}{2}g(\bar{s}_{n})f(\bar{s}_{n}) \frac{1}{\hbar^2} \norm{\big[\hat{H}_{\mathrm{fin}} ,\hat{H}_{\mathrm{ini}} \big]} \ll \frac{\tau \delta s_{n}}{\hbar} \norm{f(\bar{s}_{n}) \hat{H}_{\mathrm{ini}} + g(\bar{s}_{n}) \hat{H}_{\mathrm{fin}}} ,
\end{equation}
then we can satisfy the approximation given in Eq.~\eqref{ApEq:Approximation}. Then, a \textit{sufficient} (but not necessary) condition to get good fidelity with a single Trotter-Suzuki decomposition of the $n$-th digitized block reads
\begin{equation}
\delta s_{n} \ll \hbar\frac{2 \norm{f(\bar{s}_{n}) \hat{H}_{\mathrm{ini}} + g(\bar{s}_{n}) \hat{H}_{\mathrm{fin}}} }{\tau g(\bar{s}_{n})f(\bar{s}_{n})\norm{\left[\hat{H}_{\mathrm{fin}} ,\hat{H}_{\mathrm{ini}} \right]}  } ~~, \forall s_{n} \in [0,1].
\end{equation}

Therefore, following the strategy of obtaining a general and robust condition for the adiabatic regime, we first assume the continuum of values for the above equation, replacing $\bar{s}_{n} \rightarrow s$ (minimizing over $s$ is, at least, as good as minimizing over $\bar{s}_{n}$), and we also consider an estimate of a reasonable total run time by invoking the condition for the adiabatic theorem to write
\begin{align}
	\tau \sim \tau_{\mathrm{ad}} = \max_{0\leq s \leq 1} \hbar \frac{\norm{d_{s}H(s)}}{\Delta^2(s)} \label{Eq:AdCond} ,
\end{align}
\revisionRefA{where $\Delta(s) = \min_{n,m|n\neq m} |g_{n}-g_{m}|$ is the minimum energy gap between \textit{different} energy levels of the Hamiltonian spectrum. It is timely to mention that the above estimate also works for adiabatic Hamiltonians \textit{with spectral degeneracy}, as $\Delta(s)$ is always smaller or equal to $g_{nk}(s)$, for any $n$ and $k$, constructed by definition. In fact, as discussed in Refs.~\cite{Sarandy:05-1} and~\cite{Jansen:07}, because $\Delta(s)$ is computed for different energy levels $n$ and $m$, we can deal with degenerate Hamiltonian because the possibility of zero-gap due to two degenerate states is taken into account in the theory.}

Since the above condition needs to be satisfied for all $s_{n}$ in the digitized time domain, we can write it in a short way as
\begin{equation}
	\delta s_{n} \ll \min_{0\leq s \leq 1} \left[\frac{2\Delta^2(s)\norm{f(s) \hat{H}_{\mathrm{ini}} + g(s) \hat{H}_{\mathrm{fin}}} }{\norm{d_{s}H(s)} g(s)f(s) \norm{\big[\hat{H}_{\mathrm{fin}} ,\hat{H}_{\mathrm{ini}} \big]} }\right] .
\end{equation}

\revisionRefC{Therefore, we state the following theorem (see~\cite{SM} for further details on the theorem and its immediate consequences).
	\begin{theorem}[ST-DAT] \label{Theorem}
		Given an adiabatic Hamiltonian $\hat{H}(s) = f(s) \hat{H}_{\mathrm{ini}} + g(s) \hat{H}_{\mathrm{fin}}$, a sufficient condition for DAT decomposition into first-order of the Suzuki-Trotter decomposition,
		\begin{equation}
			\hat{U}_{\mathrm{d}}(s_{n+1};s_{n}) \approx
			e^{- \frac{i\tau \delta s_{n}}{\hbar}g(\bar{s}_{n}) \hat{H}_{\mathrm{fin}} }e^{- \frac{i\tau \delta s_{n}}{\hbar}f(\bar{s}_{n}) \hat{H}_{\mathrm{ini}} } , \label{MethodsEq:Approximation}
		\end{equation}
		is given by
		\begin{equation}
			\delta s_{n} \ll \min_{0\leq s \leq 1} \left[\frac{2\Delta^2(s)\norm{\hat{H}(s)} }{ g(s)f(s) \norm{d_{s}\hat{H}(s)} \norm{\big[\hat{H}_{\mathrm{fin}} ,\hat{H}_{\mathrm{ini}} \big]} }\right]  ,
		\end{equation}
		where $\vert\vert \hat{A} \vert\vert = \sqrt{\mathrm{tr}(\hat{A}^{\dagger}\hat{A})}$ is the Hilbert-Schmidt norm, and $\Delta(s)$ is the minimum instantaneous non-vanishing energy gap of $\hat{H}(s)$.
	\end{theorem}
}

The consequences of the digitized adiabatic theorem, as introduced in the main text, are: i) It imposes a minimum value of the number of blocks $M$ for high-fidelity digitization. In fact, as $s = \sum_{n=1}^{M} \delta s_{n}$ the above equation can also be used to determine $M$; ii) The longer the adiabatic time $\tau_{\mathrm{ad}}$, the bigger the number of blocks demanded for DAEs, establishing then a connection with validity conditions for adiabatic theorem in analog evolution; iii) The approximated behavior of the adiabatic solution allows us to reduce the number of blocks by adequately managing the adiabaticity constraint over the total evolution time $\tau_{\mathrm{ad}}$. iv) The circuit complexity, given the available native gates of our device, is given by the quantum circuit that implements the interaction terms of $\hat{H}_{\mathrm{fin}}$, since no optimization in the space of parameters of the circuit is required. On the other hand, the performance of the digitized adiabatic evolution may be optimized by finding the best functions $f(s)$ and $g(s)$, which are determined by quantum adiabatic brachistochrone trajectories and responsible for driving the system from the initial ($\hat{H}_{\mathrm{ini}}$) and problem Hamiltonians ($\hat{H}_{\mathrm{fin}}$) at a short time interval $\tau \sim \tau_{\mathrm{ad}}$.

\revisionRefA{The above discussion also works even when the adiabatic Hamiltonian admits the existence of degenerated energy levels. To take these cases into account, let us first briefly discuss how to get the Eq.~\eqref{Eq:AdCond}. For a more rigorous discussion, see Ref.~\cite{SM}. To this end, we recall the validity conditions for \textit{non-degenerated} energy spectrum with~\cite{Sarandy:04}
\begin{align}
\tau \gg \tau_{\mathrm{ad}}^{\mathrm{n-deg}} = \max_{n,k|n\neq k} \max_{0\leq s \leq 1} \left(  \frac{\hbar|\bra{E_{n}(s)}d_{s}H(s)\ket{E_{k}(s)}|}{g_{nk}^2(s)}\right) , \label{Eq:Non-degCond}
\end{align}
where $g_{nk}(s) = E_{k}(s) - E_{n}(s)$ is the instantaneous minimum gap associated to the two eigenstates $\ket{E_{n}(s)}$ and $\ket{E_{k}(s)}$. First, we observe that $\norm{d_{s}H(s)} \geq |\bra{E_{n}(s)}d_{s}H(s)\ket{E_{k}(s)}| $, because $|\bra{E_{n}(s)}d_{s}H(s)\ket{E_{k}(s)}|$ is only a single element of the matrix $d_{s}H(s)$, so we can super-estimate the inequality above by substituting the maximization over the off-diagonal matrix elements $|\bra{E_{n}(s)}d_{s}H(s)\ket{E_{k}(s)}| $ by the norm of the Hamiltonian as $|\bra{E_{n}(s)}d_{s}H(s)\ket{E_{k}(s)}| \rightarrow \norm{d_{s}H(s)}$. To end, because $\Delta(s)$ is, by definition, the \textit{minimum energy gap}, it is always smaller or equal to $g_{nk}(s)$, for any $n$ and $k$. In this way, we estimate $\tau_{\mathrm{ad}}$ by doing $g_{nk}(s) = \Delta(s)$~\cite{Rezakhani:09}. Using these two observations in above equation, we get Eq.~\eqref{Eq:AdCond}. 

Now, the generalization of our estimate for the adiabatic time to systems with spectral \textit{degeneracy} can be done if we take start from the adiabatic condition for degenerate systems. In fact, consider the eigenvalue equation $H(s)\ket*{E_{k}^{d_{k}}(s)} = E_{k}(s)\ket*{E_{k}^{d_{k}}(s)}$, where we introduce the index $d_{k}=\{1,2,\cdots,N_{k}\}$ to denote the set of $N_{k}$ eigenstates of the degenerate subspace of energy $E_{k}(s)$. By doing that, the condition in Eq.~\eqref{Eq:Non-degCond} is modified as
\begin{align}
	\tau \gg \tau^{\mathrm{deg}}_{\mathrm{ad}} = \max_{d_{k},d_{n}} \left[\max_{n,k|n\neq k} \max_{0\leq s \leq 1} \left(  \frac{\hbar|\bra{E_{n}^{d_{n}}(s)}d_{s}H(s)\ket{E_{k}^{d_{k}}(s)}|}{g_{nk}^2(s)}\right)\right] , \label{Eq:degCond}
\end{align}
where the additional maximization $\max_{d_{k},d_{n}}$ is done over the degenerate subspaces $\{\ket*{E_{k}^{d_{k}}}\}$ and $\{\ket*{E_{n}^{d_{n}}}\}$. So, by using the same analysis as before, we can estimate the $\tau^{\mathrm{deg}}_{\mathrm{ad}}$ from Eq.~\eqref{Eq:AdCond}.}

\revisionRefB{
\subsection*{Decomposing Heisenberg Interaction in fundamental gates}

As a fundamental part of the adiabatic digitizing considered in this work, we show now how to get the circuit to simulate pair-wise interactions of the Hamiltonian $\hat{H}_{Tree}$. See Ref.~\cite{Nielsen:Book} for more details on how to simulate Hamiltonian evolutions using fundamental quantum gates. For the particular case of interest to our work, let us write the Hamiltonian in Eq.~\eqref{Eq:H_bethe} as
\begin{equation}
	\hat{H}_{\mathrm{Tree}} = \hbar \sum\nolimits_{\langle n, k \rangle} J_0 \hat{h}_{n, k} , \label{Meth:Eq:H_bethe}
\end{equation}
where we have the interaction term for two arbitrary qubits $(n, k)$
\begin{equation}
	\hat{h}_{n, k} = \hbar \left(\hat{\sigma}_{n}^{+}\hat{\sigma}_{k}^{-} + \hat{\sigma}_{n}^{-}\hat{\sigma}_{k}^{+}\right) . \label{Meth:Eq:H_bethe2}
\end{equation}

As part of the approximation for digitization, we assume the system evolution operator $U(t_\ell,t_{\ell-1})$ reads
\begin{align}
\hat{U}(t_\ell,t_{\ell-1}) = \exp \left( \frac{1}{i\hbar} \hat{H}_{\mathrm{Tree}} (t_\ell,t_{\ell-1})  \right) \approx \prod\nolimits_{\langle n, k \rangle} \exp \left( -i \hat{h}_{n, k} \phi_{\ell}  \right) ,
\end{align}
where we define the dimensionless parameter (angle) $\phi_{\ell} = J_0(t_\ell,t_{\ell-1})$. Therefore, we just need to show how to implement the two-qubit unitary given by $U_{n, k}(\phi_{\ell}) = \exp \left( -i \hat{h}_{n, k} \phi_{\ell}  \right)$. At this point, because $\hat{h}_{n, k}$ can be analytically diagonalized we find
\begin{align}
U_{n, k}(\phi_{\ell}) = \left(
\begin{matrix}
	1 & 0 & 0 & 0 \\
	0 & \cos (\phi ) & -i \sin (\phi ) & 0 \\
	0 & -i \sin (\phi ) & \cos (\phi ) & 0 \\
	0 & 0 & 0 & 1 \\
\end{matrix}
\right) .
\end{align}

From this equation we observe that the structure of the operator $U_{n, k}(\phi_{\ell})$ is similar to the structure of a $i\mathrm{SWAP}$ gate 
\begin{align}
	i\mathrm{SWAP} = \left(
	\begin{matrix}
		1 & 0 & 0 & 0 \\
		0 & 1 & i  & 0 \\
		0 & i &1 & 0 \\
		0 & 0 & 0 & 1 \\
	\end{matrix}
	\right) .
\end{align}

So, it means that $U_{n, k}(\phi_{\ell})$ can be efficiently simulated using $i\mathrm{SWAP}$ as the two-qubit gates. In fact, this operator we can observe that we can decompose it as single qubit rotations and the $i\mathrm{SWAP}$ gate as
\begin{align}
	U_{n, k}(\phi_{\ell}) =  &\left[ R^{z}_{n}\left(-\frac{\pi }{2}\right) R^{z}_{k}\left(-\frac{\pi }{2}\right)\right] \cdot \left[ R^{x}_{n}\left(\frac{\pi }{2}\right) R_{k}^{y}\left(-\frac{\pi }{2}\right)\right] \cdot i\mathrm{SWAP} \cdot \left[ R^{z}_{n}\left(\frac{\pi }{2}\right) R^{y}_{k}(\phi_{\ell} +\pi )\right] \cdot \nonumber \\ &\left[ R^{y}_{n}(\phi_{\ell} +\pi ) R^{z}_{k}\left(\frac{\pi }{2}\right)\right] \cdot i\mathrm{SWAP} \cdot  \left[ R^{y}_{n}\left(-\frac{\pi }{2}\right) R^{x}_{k}\left(-\frac{\pi }{2}\right)\right] ,
\end{align}
where $\hat{R}_{\eta}(\alpha) = e^{-i\alpha \hat{\sigma}_{\eta}/2}$. However, by using the relation between $i\mathrm{SWAP}$ and the CZ gate we get the a more efficient decomposition (less gates) as
\begin{align}
U_{n, k}(\phi_{\ell}) = \left[ \hat{R}^{z}_{n}\left(\frac{\pi }{2}\right)  \hat{R}^{y}_{k}\left(-\frac{\pi }{2}\right)\right] . \left[ \hat{R}^{y}_{n}\left(-\frac{\pi }{2}\right)\right]\cdot\text{CZ}\cdot \left[ \hat{R}^{y}_{n}(\phi ) \hat{R}^{y}_{k}(-\phi )\right]\cdot\text{CZ}\cdot \left[ \hat{R}^{y}_{n}\left(\frac{\pi }{2}\right)\right]. \left[ \hat{R}^{z}_{n}\left(-\frac{\pi }{2}\right) \hat{R}^{y}_{k}\left(\frac{\pi }{2}\right)\right] .
\end{align}
}

\subsection*{Correcting noise effects on entanglement generation}

Due to noise effects during the experimental implementation of the randomized measurements, the output data needs to be properly treated in order to provide the correct output entanglement entropy. We first consider that the noise causes a uniform entropy growing over the system as a whole. In this way, the noise effects correction is done by measuring the entropy of the whole system $S(\rho)$, and then subtracting it from the original experimental entropy (including the uniform system noise). So, consider that at a given time $t$ one gets the original experimental data $S(\rho)$, then the uniform noise rate contributes as $R=S(\rho)/N$, where $N$ is the total number of qubits i.e., the entropy roof value. Now if we measure the entropy of the subsystem $A$ and obtain $S(\rho_A)$, and the number of the subsystem is $N_A$, we should correct it according to the \textit{post-processed} entanglement entropy
\begin{equation}
	S_{\mathrm{post}}(\rho_A) = S(\rho_A) -R\times N_A = S(\rho_A) -\frac{S(\rho) N_A}{N} .
\end{equation}

\subsection*{\large Data and Code Availability} Both the data and numerical codes that support the plots within this paper and other findings of this study are available and they can be provided by the corresponding authors upon reasonable request. 
\subsection*{\large Acknowledgments}
We gratefully acknowlege useful discussions with Z. Liu and F. Yan. 
	This work was supported by  the National
	Natural Science Foundation of China (11934010, 12205137, 12004167), the Key-Area Research and Development Program of Guangdong Province (Grants No. 2018B030326001 and 2020B0303030001), the China	Postdoctoral Science Foundation (Grants No. 2020M671861 and 2021T140648), the Guangdong
	Provincial Key Laboratory (Grant No. 2019B121203002), Technology and Innovation Commission of Shenzhen Municipality (KQTD20210811090049034),
	the Innovation Program for Quantum Science and Technology (Grants No. ZD0301703 and ZD0102040201). K.P. and N.T.Z. acknowledge funding from The Independent Research Fund Denmark DFF-FNU. A.C.S. is supported by the European Union's	Horizon 2020 FET-Open project SuperQuLAN (899354) and Comunidad de Madrid Sinérgicos 2020 project NanoQuCo-CM (Y2020/TCS-6545). A.C.S. acknowledges the partial financial support by the São Paulo Research Foundation (FAPESP) (Grant No. 2019/22685-1).

\subsection*{\large{Authors Contributions}} 
D.T. and D.Y. supervised the project. C.-K.H. designed the devices and conducted the measurements with assistance from G.X., J.C., C.L, R.Y. H.Y. and S.Y..
C.-K.H. and D.T. conceived and designed the experiment. Y.Z. fabricated the devices supervised by S.L.. 
A.C.S., K.P. and N.T.Z. developed the theory, and 
A.C.S. and K.P. did the theoretical simulations to fit the experimental data. A.C.S. supervised the project from the theoretical aspect.
A.C.S., C.-K.H. and D.T. wrote the manuscript with feedback from all the authors.

\subsection*{\large Competing interests} \noindent The authors declare no competing interests.

\subsection*{\large Additional Information}

\noindent\textbf{Supplementary information} for this paper is available.

\noindent\textbf{Correspondence and requests for materials} should be addressed to S.L., D.T., A.C.S. or D.Y..

\newpage

%

\begin{figure}[t!]
	\includegraphics[width=\columnwidth]{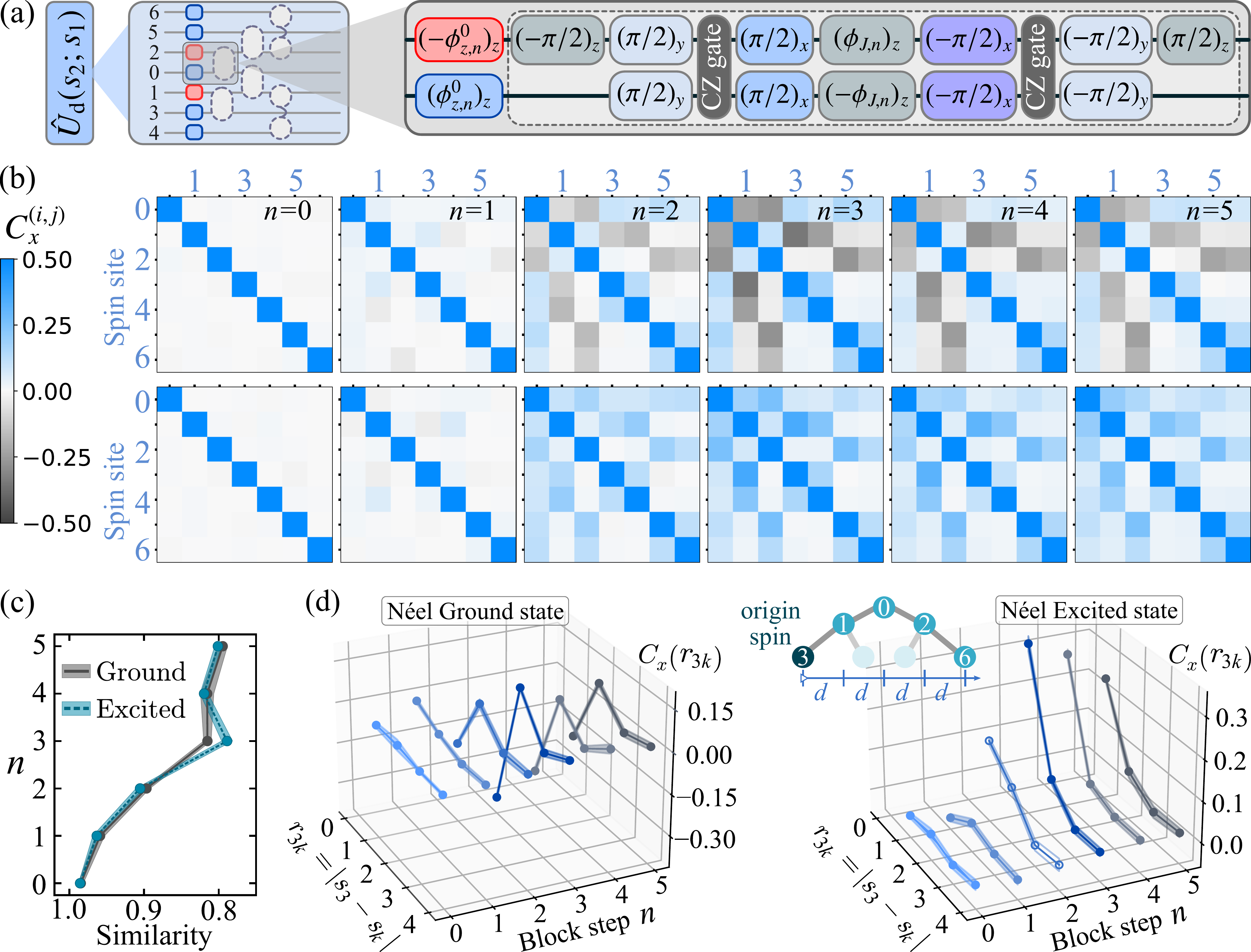}
	\caption{\textbf{Digitized circuit, correlation function $C_{x}^{(i,j)}$ and circuit fidelity.} 
		(a) Gate sequence for each block of the digitized circuit, describing how to encode the arbitrary parameters $\phi^{0}_{z,n}$ and $\varphi_{J,n}$ in the circuit to implement digitized CSB. Here we use the notation $(\alpha)_{\eta} = \hat{R}_{\eta}(\alpha) = e^{-i\alpha \hat{\sigma}_{\eta}/2}$.
		In (b) we show the experimental data of $C_{x}^{(i,j)}$ immediately after the $n$-th block of a 5-block digitized circuit. The graphs are ordered from the state initialization $(n=0)$ to the final state $(n=M=5)$, respectively, showing the digitized evolution for the system initialized in (top) the ground state and (bottom) the highest excited state. 
		Panel (c) shows the behavior of the similarity with respect to the ideal digital process, obtained for each corresponding $C_{x}^{(i,j)}$ shown in the panel (b). In (d) we present the profile of the range of $C_{x}^{3,k}=C_{x}(r_{3k})$ as a function of the Manhattan distance from the $k$-th spin to spin 3, and the block step, for the ground and excited states.}
	\label{Fig:CSB}
\end{figure}

\end{document}


\title{\Title}

\author{Chang-Kang Hu}
\affiliation{\iqa}\affiliation{\siqse}\affiliation{\gdpkl}

\author{Guixu Xie}
\affiliation{\iqa}\affiliation{\siqse}\affiliation{\gdpkl}

\author{Kasper Poulsen}
\affiliation{\DFAU}

\author{Yuxuan Zhou}
\affiliation{\iqa}\affiliation{\siqse}\affiliation{\gdpkl}

\author{Ji Chu	}
\affiliation{\iqa}\affiliation{\siqse}\affiliation{\gdpkl}

\author{Chilong Liu}
\affiliation{\iqa}\affiliation{\siqse}\affiliation{\gdpkl}

\author{Ruiyang Zhou}
\affiliation{\iqa}\affiliation{\siqse}\affiliation{\gdpkl}

\author{Haolan Yuan}
\affiliation{\iqa}\affiliation{\siqse}\affiliation{\gdpkl}

\author{Yuecheng Shen}
\affiliation{\SYU}

\author{Song Liu}
\email{lius3@sustech.edu.cn}
\affiliation{\iqa}\affiliation{\siqse}\affiliation{\gdpkl}\affiliation{\szkl}

\author{Nikolaj T. Zinner}
\affiliation{\DFAU}

\author{Dian Tan}
\email{tand@sustech.edu.cn}
\affiliation{\iqa}\affiliation{\siqse}\affiliation{\gdpkl}

\author{Alan C. Santos~\orcidlink{0000-0002-6989-7958}}
\email{ac\_santos@iff.csic.es}
\affiliation{\CSIC}
\affiliation{\UFSCar}

\author{Dapeng Yu}
\email{yudp@sustech.edu.cn}
\affiliation{\iqa}\affiliation{\siqse}\affiliation{\gdpkl}\affiliation{\szkl}\affiliation{\physustech}

\maketitle

\appendix

\setcounter{equation}{0}
\setcounter{figure}{0}
\setcounter{table}{0}

\renewcommand{\theequation}{S\arabic{equation}}
\renewcommand{\thefigure}{S\arabic{figure}}
\renewcommand{\bibnumfmt}[1]{[S#1]}
\renewcommand{\citenumfont}[1]{S#1}

\tableofcontents
\section{Device and experimental setup}

\subsection{Device and wiring}

Our tree-like superconducting quantum processor incorporates  flip-chip packaging technology. On the top layer, there are only qubits and couplers, while the bottom layer contains all control lines, transmission lines, and readout resonators, all built by coating aluminum on a sapphire substrate. A separation of approximately 9 micrometers is maintained between the bottom and top substrates.
The processor employs an architecture that combines fixed-frequency qubits with frequency-adjustable couplers ~\cite{yan2018tunable, hu2023native}. High-frequency microwave signals drive the fixed-frequency qubits to achieve high-fidelity single-qubit gates, while low-frequency signals control the couplers' frequency for high-fidelity adiabatic CZ gates~\cite{xu2020high}. Leveraging these features, qubits and their neighboring couplers can share a standard control line, reducing the overall number of control lines.

\begin{figure}[htbp]
	\centering
	\includegraphics[width=160mm]{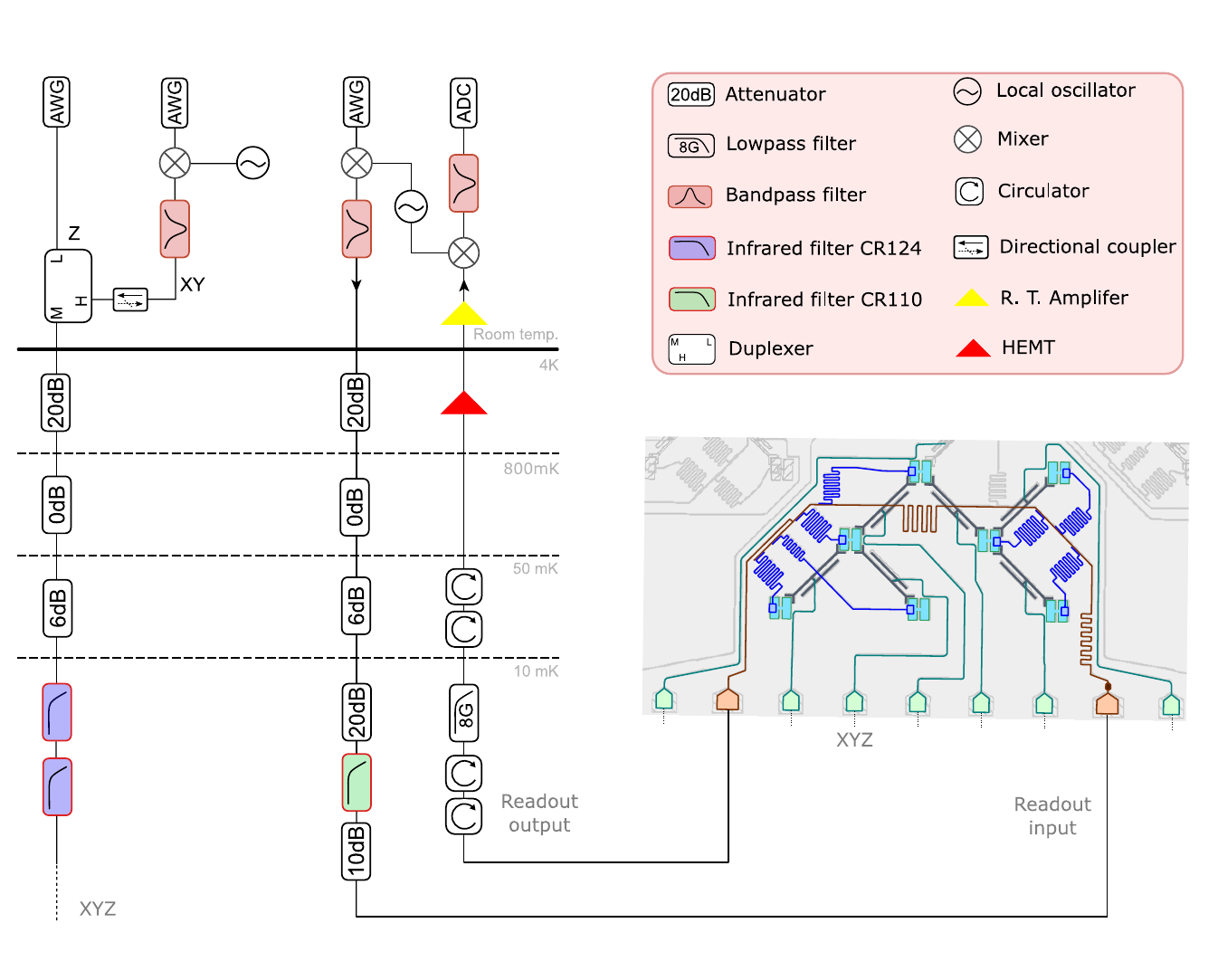}
	\caption{
		\textbf{Electronics and quantum processor schematic diagram of the experimental setup}. (Left) The schematic diagram of room temperature control electronics and wiring. (Right) The schematic diagram illustrates the tree-like superconducting quantum processor, which is a component of the complete chip. Our experimental setup only utilized a sole transmission line and its associated cavities and corresponding qubits. The depiction does not include the sections that remained unutilized due to their lack of relevance to the experiment.}
	\label{fig:wiring}
\end{figure}

We positioned the quantum processor inside a BlueFors LD dilution refrigerator operating at a base temperature of 10 mK. Additionally, we encased it within dual layers of magnetic shielding enclosures to effectively mitigate the residual magnetic field influence. As shown in Fig.~\ref{fig:wiring}, we generate the XY control waveforms required for our experiment's single-qubit gate operations by mixing the local oscillator (LO) output with the microwave pulses generated by a programmable arbitrary waveform generator (AWG). We employ a customized bandpass filter with 600 MHz bandwidth to ensure signal integrity, eliminating residual LO signals and their mirrored counterparts, yielding clean control signals. Then, we employ duplexers to merge the XY and Z signals. After filtering through attenuators and dual infrared filters, the combined signals will be transmitted to the XYZ multiplexed control lines on the chip. We generate dispersive readout signals using the same method as the XY signals. Subsequently, a high electron-mobility transistor (HEMT) amplifier at the 4K stage amplifies the readout output signal. After passing through the HEMT amplifier and a low-noise amplifier at room temperature, the readout signal is down-converted, and the resulting demodulated signals are digitized by analog-to-digital converters (ADC).

\begin{table}[h!]
	\begin{center}
		\begin{threeparttable}
			\begin{tabular}{|p{4.5cm}<{\centering}|p{1.5cm}<{\centering}|p{1.5cm}<{\centering}|p{1.5cm}<{\centering}|p{1.5cm}<{\centering}|p{1.5cm}<{\centering}|p{1.5cm}<{\centering}|p{1.5cm}<{\centering}|p{1.5cm}<{\centering}|}
				
				\hline
				\hline
				Qubit$^{\rm a}$                                                     & $Q_0$ & $Q_1$ & $Q_2$ & $Q_3$  & $Q_4$  &  $Q_5$  &  $Q_6$  \\
				\hline
				Frequency (GHz)                                                     & 4.234 & 4.620 & 4.675 & 4.099  & 4.153  & 4.027   & 4.213   \\
				Anharmonicity (MHz)                                                 & -218  & -208  & -204  & -217   & -218   & -220    & -231    \\
				Resonator frequency (GHz)                                           & 6.455 & 6.674 & 6.421 & 6.550  & 6.376  & 6.517   & 6.729   \\
				Resonator linewidth (MHz)                                           & 0.30  & 0.17  & 0.76  & 0.41   & 1.23   & 0.29    & 0.23    \\
				Dispersive shift of $|1\rangle$ (MHz)                               & 1.34  & 1.20  & 1.30  & 1.24   & 1.06   & 1.16    & 1.24    \\
				Readout fidelity of $|0\rangle$ (\%)                                & 99.0  & 97.6  & 97.8  & 98.3   & 98.7   & 97.2    & 97.8    \\
				Readout fidelity of $|1\rangle$ (\%)                                &  95.1 &  94.4 & 93.0  & 93.8   &  95.4  &  91.5   &  94.4   \\
				Relaxation time of $|1\rangle$  ($\mu s$)                           & 75.5  & 50.9  & 42.9  & 67.9   & 35.9   & 28.2    & 61.5    \\
				Ramsey decay time (isola.)($\mu s$)                                 & 32.4  & 56.1  & 28.1  & 37.1   & 25.6   & 42.6    & 32.3    \\
				Spin echo decay time (isola.) ($\mu s$)                             & 40.3  & 46.8  & 33.8  & 39.6   & 52.0   & 46.8    & 40.3    \\
				Ramsey decay time (simul.) ($\mu s$)                                & 16.6  & 4.9   & 6.1   & 7.3    & 14.7   & 17.4    & 16.6    \\
				Spin echo decay time (simul.)($\mu s$)                              & 31.9  & 5.4   & 4.6   & 28.5   & 45.4   & 11.2    & 31.9    \\
				1-Q gate error$^{\rm b}$ (simul.)(\%)                               & 0.13  & 0.05  & 0.07  & 0.12   & 0.06   & 0.06    & 0.07    \\
				
				\hline
				\hline
				CZ gate                              &  $CZ_{01}$ &  $CZ_{02}$   & $CZ_{13}$   & $CZ_{14}$ &  $CZ_{25} $ &   $CZ_{26}$ &   $-$\\
				\hline
				CZ gate error$^{\rm c}$ (simul.)(\%) & 1.42       &  1.37        & 0.90        & 1.12      & 2.09      & 1.02          & -\\
				\hline
				\hline		
				
			\end{tabular}
			\begin{tablenotes}
				\footnotesize
				\item[a] These parameters are measured while the coupler is idle at the nearest ZZ coupling closed point. 
				\item[b] We simultaneously implement the cross-entropy benchmarking (XEB) to test the single-qubit gates' fidelity and give the final Pauli error according to the decay rate.
				\item[c] The CZ gate errors are measured with XEB. The simultaneous two-qubit XEB is performed by executing a standard XEB in the target two-qubit gate and adding randomized single-qubit gates in the irrelevant qubits. The single-qubit gates are aligned in quantum circuits.
			\end{tablenotes}
		\end{threeparttable}
	\end{center}
	\caption{\textbf{Device parameters}.}
	\label{table:Device}
\end{table}	

\subsection{Device parameters}

We have compiled a comprehensive list of the primary system parameters, as presented in the Table~\ref{table:Device}. These parameters encompass but are not restricted to qubit frequency, $T_1$, $T_2$, read fidelity, single-qubit gate fidelity, two-qubit gate fidelity, and others.

In our experimental setup, the average $T_1$ for the employed qubits extended to 51.9 $\mu s$, and the average $T_2$ reached 9.9 $\mu s$. Concerning the quantum state readout, even in the absence of Josephson Parametric Amplifier (JPA), we have achieved an average fidelity of 96.0\%  across all measurements. The cross-entropy benchmarking (XEB) experiments show that the simultaneously single-qubit gates are executed with an  average fidelity of 99.92\% , closely approaching the $T_1$ limit, and the two-qubit gates maintained an average fidelity of 98.68\%. We hypothesize that the fidelity of two-qubit gates in the XEB protocol is primarily limited by the qubits' $T_2$ and stray couplings within the chip.

\revisionRefC{
Dissipation, which can prematurely destroy quantum states, is a significant challenge in quantum computing. In our system, the primary sources of dissipation are environmental coupling (e.g., quantum noise, electromagnetic radiation, material defects) 
, device noise (e.g., electronic noise, control noise), and operational errors (e.g., gate errors, measurement errors). These dissipative processes contribute to quantum relaxation and decoherence, thereby limiting gate fidelity. To quantify their impact, we characterize the system based on the relaxation times, dephasing times, and gate fidelity presented in Table~\ref{table:Device}.
}

\subsection{Device effective temperature}

\begin{table}[b]
	\centering
	\caption{Thermal population and effective temperatures}
	\begin{tabular}{|c|c|c|c|c|c|c|c|c|c|}
		\hline
		& $Q_{0}$ &  $Q_{1}$ & $Q_{2}$ &  $Q_{3}$ &  $Q_{4}$ &  $Q_{5}$ &  $Q_{6}$ & Mean & Std. Dev.  \\
		\hline
		Thermal population in $\ket{1}$ ($\%$) & 1.0 & 2.4& 2.2& 1.7& 1.3& 2.8& 2.2& 1.94 & 0.59 \\
		\hline
		Effective temperatures (mK) & 44.2&59.8&59.1&48.4&46.0&54.4&53.2 & 52.15 & 5.71 \\
		\hline
	\end{tabular}
	\label{table:thermal}
\end{table}

\revisionRefB{
Differently from the target zero-temperature dynamics considered in the digital circuit implemented in our work, the quantum hardware is not at absolute zero temperature. This can lead to different error sources: quantum state initialization, quantum circuit operation, and quantum state measurement. These errors introduce a small but non-zero probability of the system being excited to higher energy states, which, however, has a negligible impact on our main conclusions. 

Regarding quantum state initialization, we prepare our qubits by damping them to thermal equilibrium. Due to the limitations of the dilution refrigeration system, the base temperature is approximately 10~mK. Consequently, our initial state is a thermal state characterized by a Boltzmann distribution with an average thermal population in $\ket{1}$ of 1.94$\%$ and effective temperature of 52.15~mK, rather than a pure ground state. The estimate (effective) temperature of each individual qubit in the initial state preparation is shown in Table \ref{table:thermal}.
}

\subsection{Crosstalk}
\begin{figure}[t!]
	\centering
	\includegraphics[width=165mm]{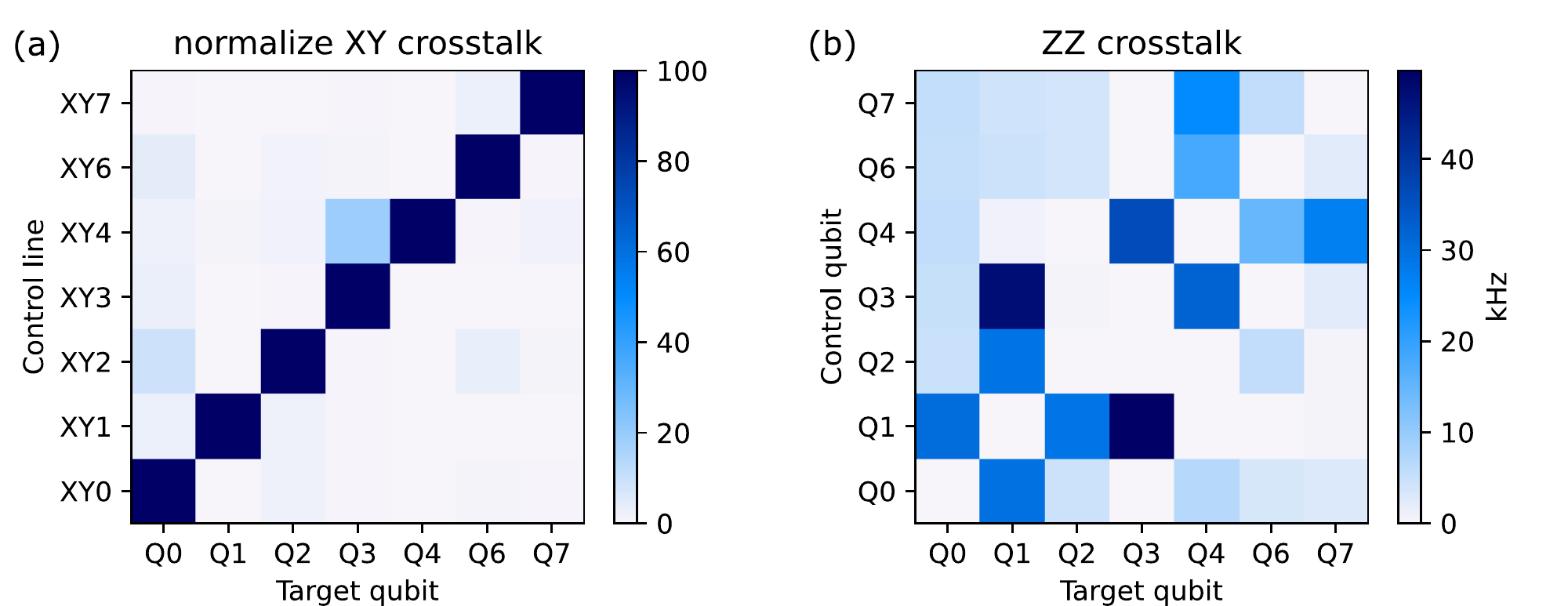}
	\caption{
		\textbf{Normalize XY crosstalk matrix and ZZ crosstalk matrix}. (a)The normalize XY crosstalk matrix. (b)The ZZ interaction strength is measured with all the couplers biased at the idling point.}
	\label{fig:xyzcrosstalk}
\end{figure}

Crosstalk is a primary obstacle in realizing superconducting quantum computing. As illustrated in Fig.~\ref{fig:xyzcrosstalk}, we have showcased the normalized XY and ZZ crosstalk. Notably, Fig.~\SubFig{fig:xyzcrosstalk}{a} shows the XY crosstalk remains 1.5\% on average. We have minimized XY crosstalk by optimizing the sample's design and fabrication phases with the techniques that incorporated a coverage bridge, entailing an additional layer of aluminum overlaying the XY control lines. It effectively confines XY signals within the transmission lines, leading to a substantial reduction in signal interference. The primary source of XY crosstalk is localized at the ends of the XY lines and the wire bonding.

Fig.~\SubFig{fig:xyzcrosstalk}{b} shows that the graph illustrates the ZZ crosstalk values obtained after calibration with the coupler at its idling point. It is noticeable that, even though complete elimination of all ZZ couplings is not achievable, they are relatively modest in magnitude and predominantly localized between nearest-neighbor qubits. Among the qubits we utilized, the maximum ZZ coupling observed stands at 47.3 kHz. As illustrated in Table \ref{table:Device}, the minimized cross-talk guarantees we can achieve fast, high-fidelity parallel single-qubit gates.

\subsection{Readout crosstalk and correction}

\begin{figure}[htbp]
	\centering
	\includegraphics[width=130mm]{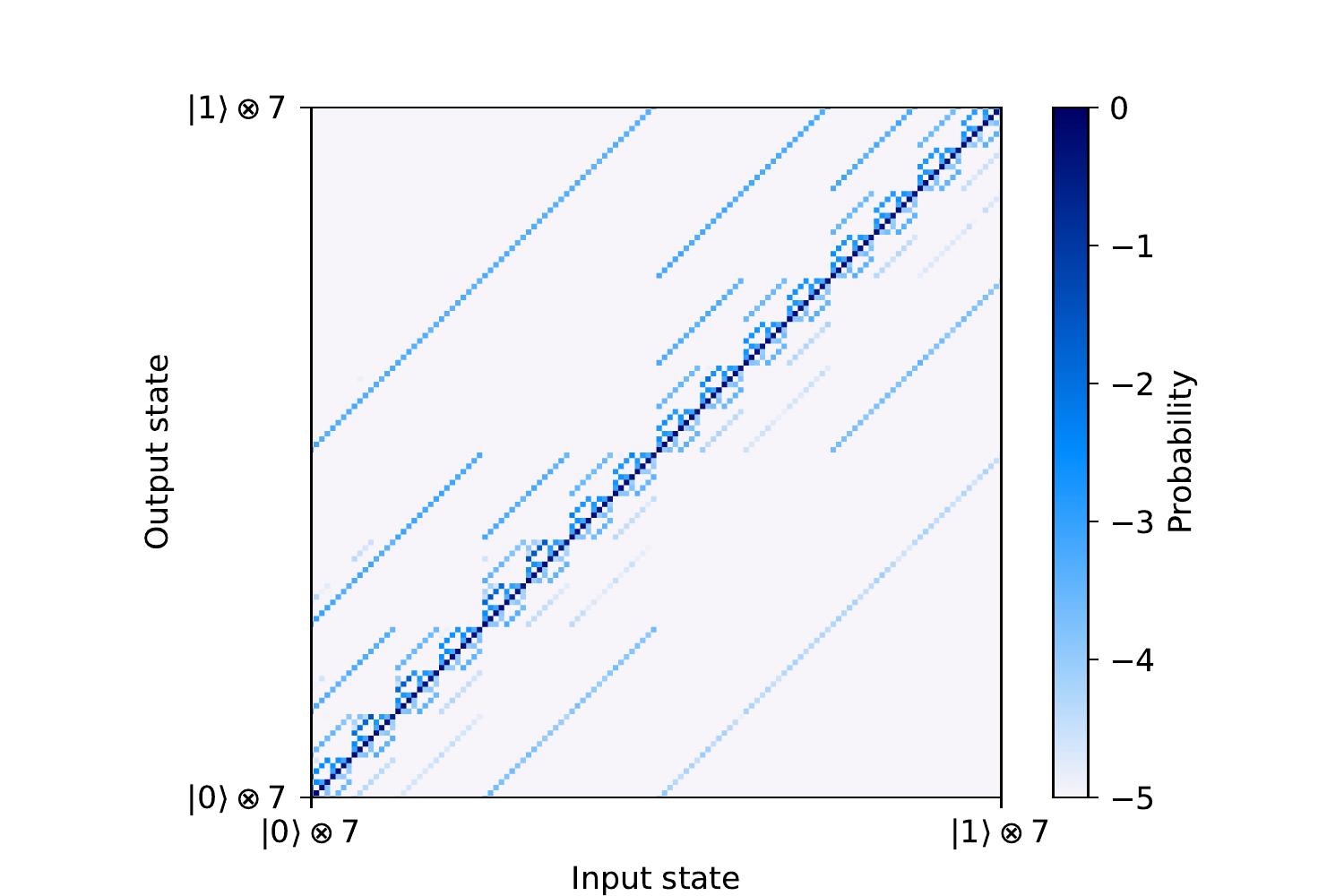}
	\caption{
		\textbf{Multi-qubit readout crosstalk matrix}. Our experiment systematically generated all $2^7 = 128$ possible quantum states and conducted total joint measurements for each prepared state to obtain the associated probabilities of states in various bases. The graph shows that the horizontal axis represents our input states, while the vertical axis displays our measurement results. For each prepared state, we repeat the binary measurement for 10000 times. To improve clarity, we have applied a logarithmic scale to the probabilities.}
	\label{fig:readcrosstalk}
\end{figure}

The errors encountered during the readout originate from two primary sources. The first type of error pertains to mapping inaccuracies involving misjudgments between the 0 and 1 states, as the device parameters Table \ref{table:Device} shows that our average readout fidelity is 96\%. The second type of error results from readout crosstalk, where changes in the quantum states of neighboring qubits may influence the determination of the target qubit's state. The accompanying Fig.~\ref{fig:readcrosstalk} illustrates the readout cross-talk scenario for all the qubits utilized in our experimental setup.

Both types of errors can be mitigated by applying a suitable mapping matrix $P_{corr}= \mathcal{M}P_{expe}$ ~\cite{hu2023native}, where $P_{expe}$ is the experimental populations, $P_{corr}$ is the corrected populations, $\mathcal{M}$ is the mapping matrix which can be determined through experimental calibration. In our experiments, we apply such correction to the two target qubits  during the calculation of $C_{x}^{(i,j)}$.

\section{Analog simulation of zero-temperature SSB}  \label{Sec:ApCSB}

Here we present a direct comparison between the adiabatic evolution of SSB for a system of $N$ spins with slowly decaying flip-flop interaction and its nearest neighbor counterpart. Firstly, let us focus on the case of the linear spin chain, where the system is initially prepared in eigenstates of the staggered-field Hamiltonian
\begin{align}
	\hat{H}_{0} = \sum\nolimits_{n=0}^{N-1} \hbar (-1)^{n}\omega_0 \hat{\sigma}_{n}^{z} .\nonumber
\end{align}

As in the main text, here we focus on the lowest and highest energy states $\hat{H}_{0}$ because they are AFM Néel states. Then, the analog simulation of the dynamics is done by the adiabatic evolution driving the system from the ground (excited) AFM Néel state to the quantum AFM (FM) state of the final XY Hamiltonian. Here, the final (problem) Hamiltonian considered here is the slowly decaying interaction Hamiltonian given as
\begin{align}
\hat{H}_{\mathrm{sd}} = \sum\nolimits_{\langle i,j \rangle} \frac{J_0}{|i-j|} \left(\hat{\sigma}^{+}_{i}\hat{\sigma}^{-}_{j} + \hat{\sigma}^{-}_{i}\hat{\sigma}^{+}_{j}\right),\nonumber
\end{align}
where we use the index ``sd" denoting the slowly decaying interacting model. Also, we consider the case of the nearest-neighbor interacting case encoded in the final Hamiltonian
\begin{align}
	\hat{H}_{\mathrm{nn}} = \sum\nolimits_{i=0}^{N-1} J_0 \left(\hat{\sigma}^{+}_{i}\hat{\sigma}^{-}_{i+1} + \hat{\sigma}^{-}_{i}\hat{\sigma}^{+}_{i+1}\right).\nonumber
\end{align}

\begin{figure}[h!]
	\includegraphics[width=\columnwidth]{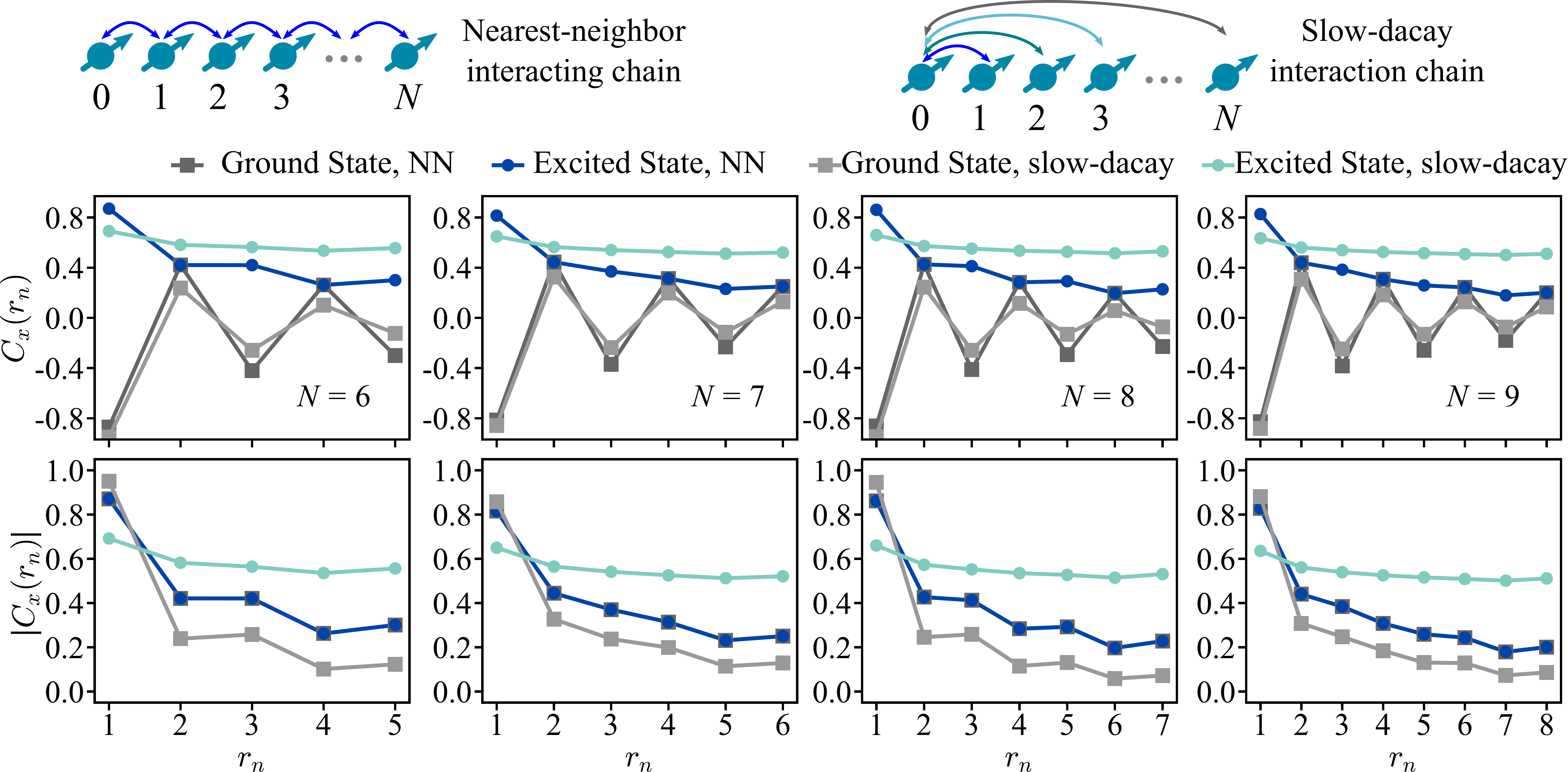}
	\caption{\textbf{Analog simulation of CSB for a linear chain.} Here we set the initial Hamiltonian parameters and the adiabatic time such that $\omega_{0} = J_0$ and $J_0\tau = 50$, which is sufficient to guarantee the adiabatic approximation for all values of $N$ considered. For simplicity of the numerical simulation, in this dynamics we use the linear interpolation for the functions $f,g$ given by $f(s)=1-s$ and $g(s)=s$.}
	\label{SM:Fig:LongRange}
\end{figure}

So, for the sake of comparison of the correlation length of both models, where we define the ``distance" as $r_{n}$ by taking into account the first spin of the chain as the reference spin, labeled with index $0$ and sketched in Fig.~\ref{SM:Fig:LongRange}, such that $r_{n} = |s_n - s_0| = n$. Then we simulate both dynamics for chains with different number of spins (from $N=6$ up to $N=9$). In such a figure we show both $C_{x}(r_{n})$, which allows us to observe the formation of FM and AFM phases depending on the initial state (signature of SSB), and its absolute value $|C_{x}(r_{n})|$, which allows us to observe the profile of the correlations as a function of the separation between two spins. From the results shown in Fig.~\ref{SM:Fig:LongRange}, we can see that an intermediate-range correlation arises in the simulation. We call it intermediate-ranged in order to distinguish it from the regimes of correlation reached for the AFM and FM phases generated for the slowly decaying interaction case. Also, from Fig.~\ref{SM:Fig:LongRange}, it is important highlighting the finite size effect observed when $r_{n}$ approaches to $r_{N-1}$. Such a behavior is observed for both kinds of interactions of the linear chain.

\begin{figure}[t!]
	\centering
	\includegraphics[width=\columnwidth]{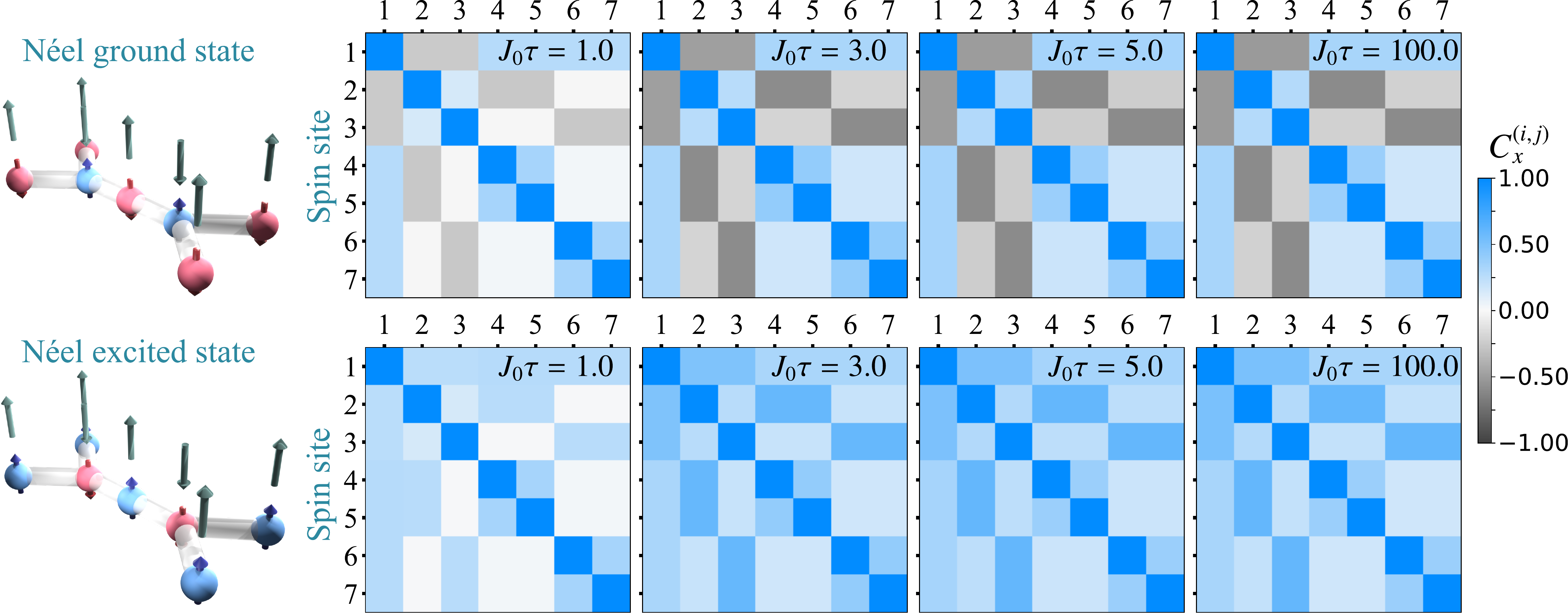}
	\caption{\textbf{CSB signature for a 7-qubit tree-like lattice.} The simulated system and the sketch of the ground and excited Néel states are shown, as well as the profile of the correlation function for each phase. We use $\omega_{0} = J_0$.}
	\label{SM:Fig:BTN-CorrelationFunctios}
\end{figure}

\begin{figure}[h!]
	\centering
	\includegraphics[width=\columnwidth]{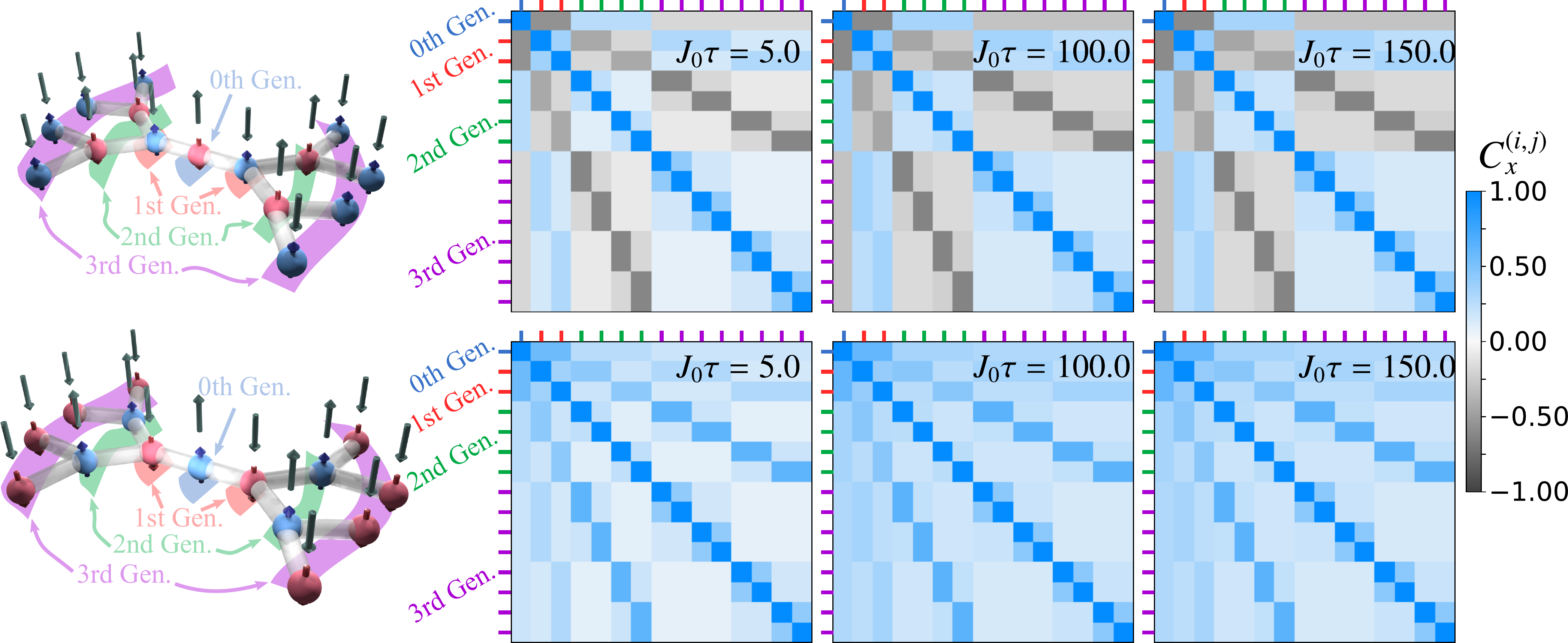}
	\caption{\textbf{CSB signature for a 15-qubit tree-like lattice.} The topology considered in the simulation and the sketch of the ground (top) and excited (bottom) Néel states are shown. Also, the full matrix for the connected correlations $C_{x}^{(i,j)}$ as a function of the different generations of the system. We use $\omega_{0} = J_0$.}
	\label{SM:Fig:BTN15Spins}
\end{figure}

Now, in order to support the statement in the main text on the correlation range for the analog version of the tree-like lattice SSB, we also present the correlations expected for the ideal case. First of all, let us show that our choice in the main text for $J_{0}\tau = 5$ is enough to observe the SSB signature in the analog evolution. Considering the time-dependent Hamiltonian given by
\begin{equation}
	\hat{H}(s) = f(s)\left[\hbar\sum\nolimits_{l=0}^{L-1}\sum\nolimits_{n_{l}=1}^{N_{l}}  (-1)^l \omega_0 \hat{\sigma}_{n_{l}}^{z}\right] + g(s) \left[\hbar \sum\nolimits_{\langle n, k \rangle} J_0 \left(\hat{\sigma}_{n}^{+}\hat{\sigma}_{k}^{-}+\hat{\sigma}_{n}^{-}\hat{\sigma}_{k}^{+}\right)\right] , \label{Eq:Ap:Hadiaba}\nonumber
\end{equation}
where $\sum_{\langle n, k \rangle}$ is a sum over all connections of the Tree lattice, as shown in Fig.~\ref{SM:Fig:BTN-CorrelationFunctios}, and where we consider $f(s)$ and $g(s)$ as given the Brachistochrone defined in the main text. To this end, we numerically solve the Schrödinger equation for four different total evolution times given by $J_{0}\tau = \{1.0, 3.0, 5.0 , 100.0\}$, where the case $J_{0}\tau = 100.0$ is definitely inside the adiabatic evolution regime. So, for each choice of $J_{0}\tau$ we extract the final state of the evolution $\rho(s=1)$ and compute the respective correlation functions $C_{x}(i,j)$, which are shown in Fig.~\ref{SM:Fig:BTN-CorrelationFunctios}. It is worth highlighting that the cases $J_{0}\tau = 5.0$ and $J_{0}\tau = 100.0$ are almost indistinguishable in these plots, so our numerical analysis suggests that $J_{0}\tau = 5.0$ is enough to observe signatures of SSB. Also, for completeness, in Fig.~\ref{SM:Fig:BTN15Spins} we present the analog expected result when include one more generation to the 7 qubits case, leading to a 15 qubits lattice, and increasing the total evolution time as $J_{0}\tau = \{5.0, 100.0, 150.0\}$. Also, it is possible to see that $J_{0}\tau = 5.0$ is enough to observe the SSB phase, when comparable with the case $J_{0}\tau=150.0$ (adiabatic regime).

\begin{figure}[h!]
	\centering
	\includegraphics[width=\linewidth]{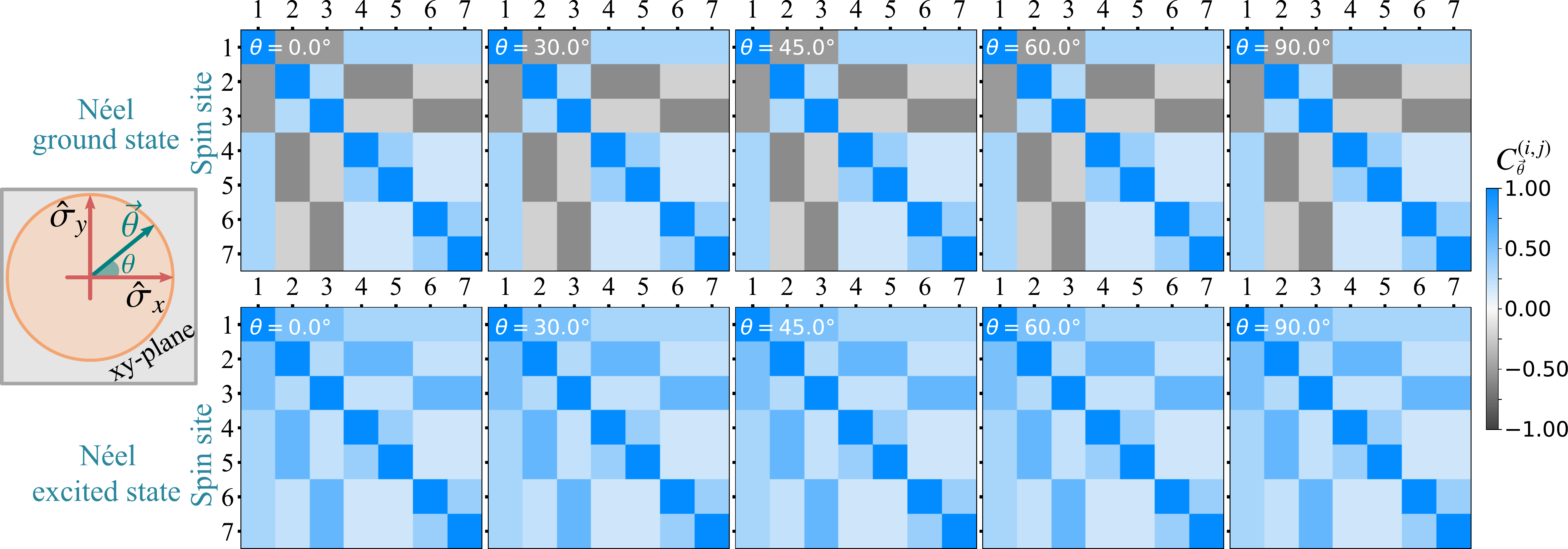}
	\caption{\textbf{Correlation function for the 7-qubit tree-like lattice.} Correlation function for the analog simulation of the adiabatic evolution of each initial state considered. Parameters as given in \ref{SM:Fig:BTN-CorrelationFunctios}.} \label{SM:Fig:CUniform}
\end{figure}

\revisionRefB{As discussed in the main text, the experimental data collected to demonstrate the SSB and phase transition to correlated phase of the matter have been done by computing the correlation function only along the $x$-direction. It is because similar results can be obtained when we measure correlations along any direction in XY plane. To observe this result, one can define the magnetization operator $\hat{\sigma}_{xy}(\theta) = \cos(\theta)\hat{\sigma}_{x} + \sin(\theta)\hat{\sigma}_{y}$, which gives us the magnetization along the direction $\vec{\theta}$ in the plane XY as shown in Fig.~\ref{SM:Fig:CUniform}. By varying $\theta$ we can observe no relevant change in the correlation profile. For example, by measuring the correlations along the y-axis ($\theta = \pi/2$) we get a similar result as correlations along the x-axis ($\theta = 0$).}

\begin{figure}[h!]
	\centering
	\includegraphics[width=0.9\columnwidth]{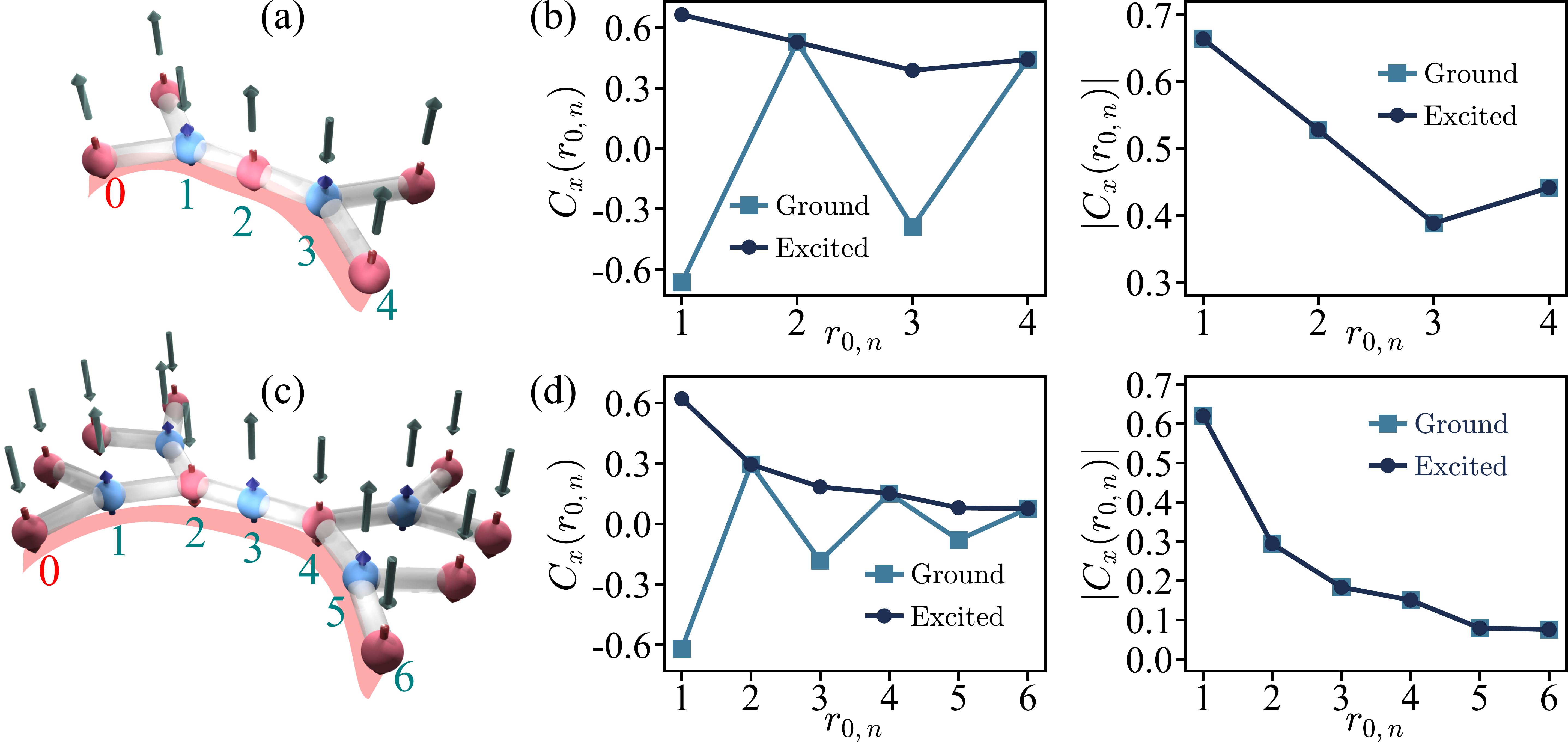}
	\caption{\textbf{Correlation length profile for the tree-like lattices.} The (a) and (c) show how we encode the ``position" of the spins to define the distance $r_{0,j}$, where the reference spin is highlighted with the index 0 in red text. In (b) and (d) we show the value of the correlation function $C_{x}(r_{0,j})$ (and its absolute value $|C_{x}(r_{0,j})|$) as function of $r_{0,j}$. As seen in Figs.~\ref{SM:Fig:BTN15Spins} and~\ref{SM:Fig:BTN-CorrelationFunctios}, due to the symmetry of the couplings, we expect that other choices of labels will provide the same result. The parameters for these results are the same as in Figs.~\ref{SM:Fig:BTN15Spins} and~\ref{SM:Fig:BTN-CorrelationFunctios}.}
	\label{SM:Fig:BTN}
\end{figure}

As a last discussion, we also give the theoretical support (by simulations) of the discussion done in the main text about the decay of correlation function $C_{x}^{(i,j)}$ with the ``distance" between $i$-th and $j$-th spins. To this end, in Fig.~\ref{SM:Fig:BTN} we show the behavior of $C_{x}(r_{0,j})$ with $r_{0,j}$, the distance between the reference spin (labeled as spin $0$) and the spin $j$. We consider here the system of 7 and 15 spins, where we also present a schematic figure showing how $r_{0,j}$ is defined. As already stated, our results indicate that the relatively low amount of correlations in the experiment is mainly due to gate errors, and it is not a characteristic of the SSB phase itself. 

\begin{figure}[h!]
	\centering
	\includegraphics[width=\columnwidth]{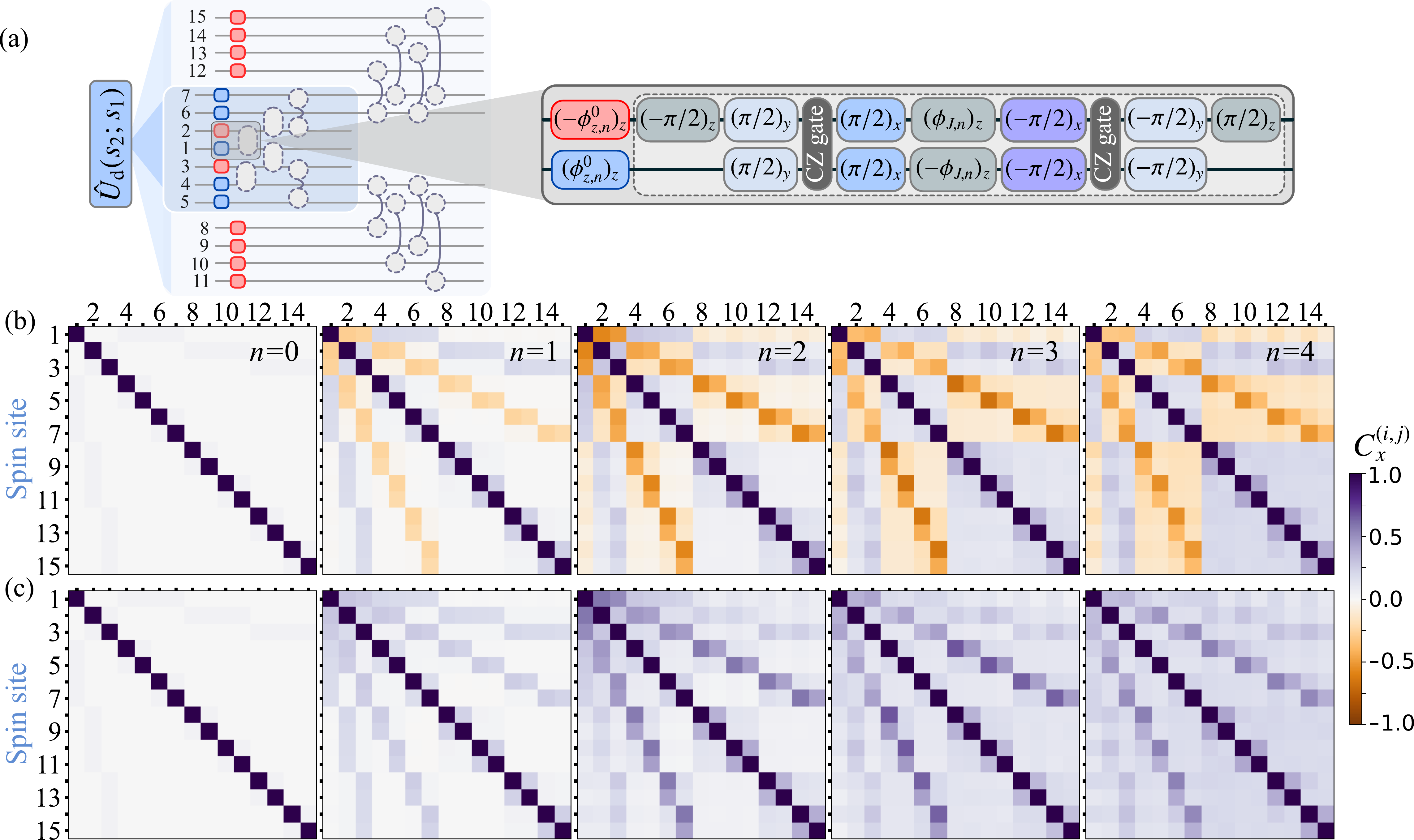}
	\caption{\textbf{Correlation function for the digitized circuit of 15 qubits.} (a) Circuit digitized evolution of a 15 qubits system, where the additional required gates are shown as a ``extension" of the 7 qubits circuit. Correlation function for the system when the dynamics starts in (b) the ground state and (c) the excited state. The parameters for these results are the same as in Fig.~\ref{SM:Fig:BTN15Spins} with $J_{0}\tau = 5$.}
	\label{SM:Fig:BTN15Digital}
\end{figure}

\revisionRefB{
\section{Digitized annealing for larger systems}  \label{Sec:ApScalling}

In this section we briefly discuss about the complexity of implementing digitized adiabatic evolutions for large systems. To this end, let us consider the Hamiltonian of the form $\hat{H}(t) = f(t) \hat{H}_{\text{ini}} + g(t) \hat{H}_{\mathrm{fin}}$. In general, adiabatic algorithms admit the implementation starting from a very easy initial Hamiltonians $\hat{H}_{\text{ini}}$ with only single-qubit local fields, like the one considered in our work. However, the main complicated part is always the final $ \hat{H}_{\mathrm{fin}}$, and sometimes referred to as \textit{the problem} Hamiltonian. So, as any other quantum algorithm or quantum optimization methods, the main complexity of our approach comes entirely from the realization of this Hamiltonian because it may contain many-body terms, long-range or all-to-all interactions. However, given this complexity, we will focus here on the scenario in which digital annealers will be advantageous.

First of all, when $\hat{H}_{\mathrm{fin}}$ can be efficiently implemented in a given processor, it is easy to conclude that the total complexity will be $C = M \times \Dcal_{C}$ where $M$ is the number of blocks and $\Dcal_{C}$ is the circuit depth for each block. Therefore, given the inevitable increasing of $\Dcal_{C}$ with the number of qubits and interactions of $\hat{H}_{\mathrm{fin}}$, we show now that the digital annealing may be benefited by the fact that $M$ may be constant for systems with different number of qubits. To this end, we will consider the immediate application of our interest, the digital simulation of the results shown in Fig.~\ref{SM:Fig:BTN15Spins} for the SSB of a 15-qubit tree-like lattice.

As shown in Fig.~\ref{SM:Fig:BTN15Digital}, the circuit for each block needs to be increased to account for all interactions in te system, but each interaction is implement with the same circuit as the 7 qubit case. So, even using a smaller number of blocks for digital circuit ($M=4$), and include 8 more qubits in the system, we can see the emergence of SSB without no significant impact for the protocol. We considered a $4$ blocks circuit because the learning acquired with the 7 qubits case, where the SSB can be observed when $n \geq 3$. Therefore, it shows that in fact the digitization is not drastically affected by increasing the system from 7 to 15 qubits.
}

\begin{figure}[h!]
	\centering
	\includegraphics[width=0.85\columnwidth]{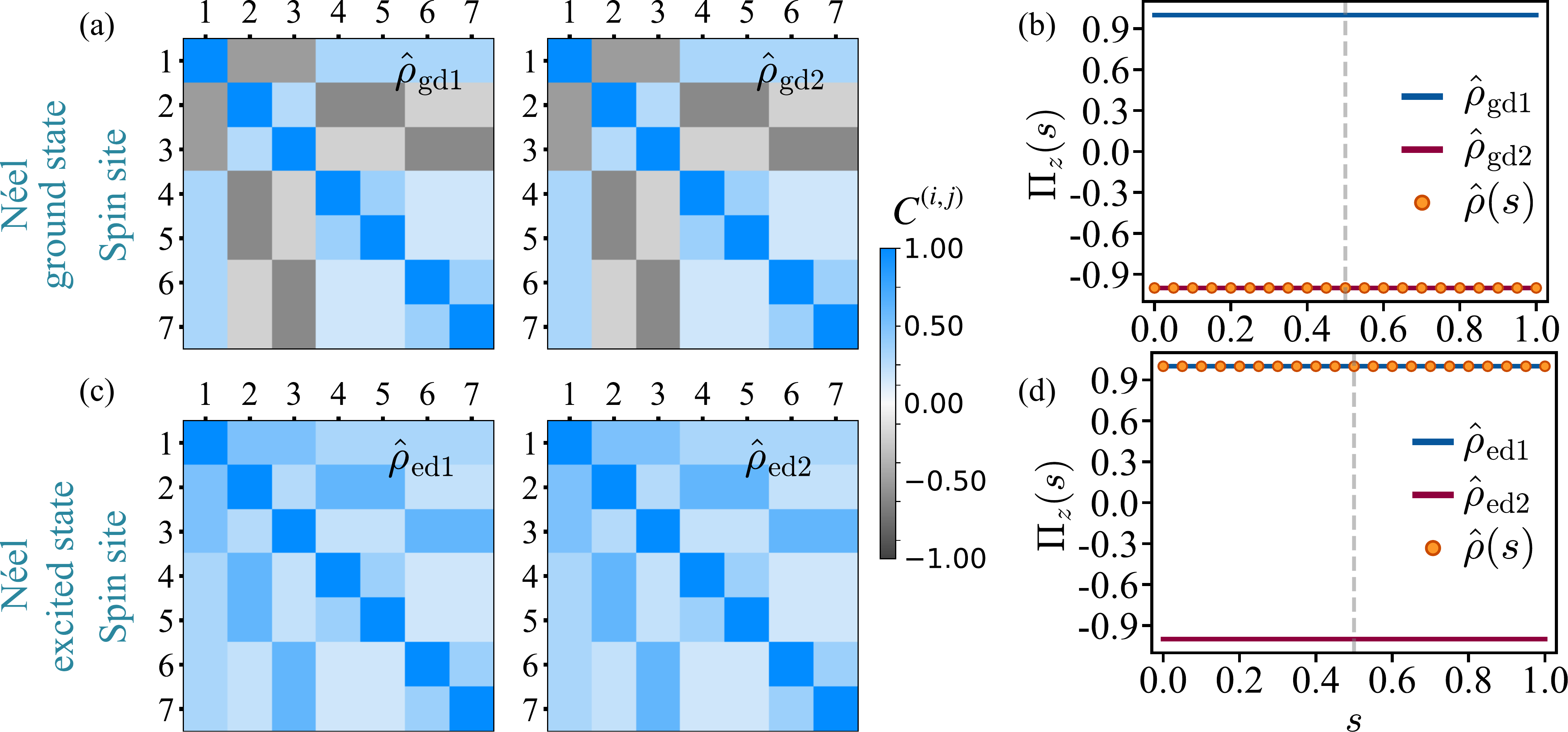}
	\caption{\textbf{Correlation profile and parity dynamics.} Panels (a) and (c) show profile of the correlation function $C_{x}^{i,j}$ for the two degenerate states of the energy level of interest for each dynamics considered. In panels (b) and (d) it is possible to see the expected value of the $Z$-parity for each degenerate state (continuum lines), and for the evolved state (circles) obtained for each initial states. The parameters for these results are the same as in Fig.~\ref{SM:Fig:BTN-CorrelationFunctios}.}
	\label{SM:Fig:Symmetry}
\end{figure}

\revisionRefB{
\section{Degeneracy of the Hamiltonian and symmetries}  \label{Sec:ApSymmetries}

Now, we will discuss more details about the properties of the adiabatic Hamiltonian $\hat{H}(s) = f(s)\hat{H}_{\text{Néel}} + g(s)\hat{H}_{\mathrm{Tree}}$ and its symmetries and degree of degeneracy.

Although the initial state of the system and its Hamiltonian are not degenerate for the ground state, the final Hamiltonian $\hat{H}_{\mathrm{Tree}}$ is degenerate due to the following symmetries,
\begin{align}
\hat{\Pi}_{\eta} = \prod_{l=0}^{L-1} \prod_{n_{l}=1}^{N_{l}} \hat{\sigma}_{n_{l}}^{\eta} = \hat{\sigma}_{1_{0}}^{\eta}\hat{\sigma}_{1_{1}}^{\eta}\hat{\sigma}_{2_{1}}^{\eta}\hat{\sigma}_{1_{2}}^{\eta}\hat{\sigma}_{1_{2}}^{\eta}\hat{\sigma}_{2_{2}}^{\eta}\hat{\sigma}_{3_{2}}^{\eta}\hat{\sigma}_{4_{2}}^{\eta} , ~~ \forall \eta \in [x,y,z]. \nonumber
\end{align}
More than that, all energy states of the Hamiltonian are (at least) doubly degenerate. Therefore, we need help of these symmetries to make sure that undesired transitions inside the same degenerate energy level are not allowed. As shown below, thanks to the parity symmetries the dynamics of the system leads to a single target state of relevant degenerate subspace.

As example, let us consider the case in which the system is initially the classical AFM Néel state. Through a simple numerical analysis of the final Hamiltonian spectrum, the degenerate subspace in which the evolution happens is composed of two states we call $\hat{\rho}_{\mathrm{dg}1}$ and $\hat{\rho}_{\mathrm{dg}2}$, which are both correlated quantum antiferromagnetic states as shown in Fig.~\SubFig{SM:Fig:Symmetry}{a}. Even under this complicated situation, we still can show how the system goes through the path to achieve the state as final state. In fact, let us now use the symmetries of the system. 

Among the possibilities aforementioned, we choose the (total) Z-parity defined as $\hat{\Pi}_{z}$. Because the adiabatic Hamiltonian $\hat{H}(s) = f(s)\hat{H}_{\text{Néel}} + g(s)\hat{H}_{\mathrm{Tree}}$ satisfies $[\hat{\Pi}_{z},\hat{H}(s)] = 0$ for any $g$ and $f$, it is expected that the evolved state conserves the initial parity of the system. In Fig.~\SubFig{SM:Fig:Symmetry}{b} we show the parity of the evolved states and the parity of the final degenerate eigenstates $\hat{\rho}_{\mathrm{dg}1}$ and $\hat{\rho}_{\mathrm{dg}2}$. As we can see, the possibility of transition of the system state to $\hat{\rho}_{\mathrm{dg}1}$ is forbidden by symmetry. In a similar way, the Figs.~\SubFig{SM:Fig:Symmetry}{c} and~\SubFig{SM:Fig:Symmetry}{d} show the same analysis for the case in which the dynamics of the system starts in the excited Néel state.}

\begin{figure}[h!]
	\centering
	\includegraphics[width=0.85\linewidth]{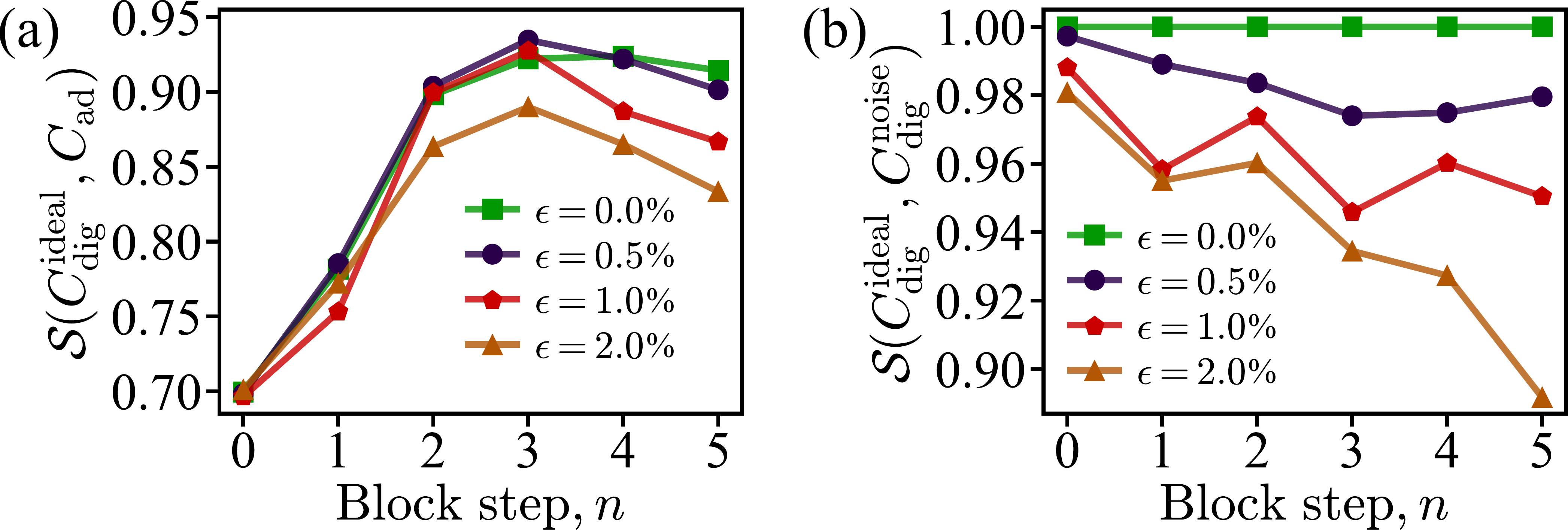}
	\caption{\textbf{Gate error analysis.} Panels (a) and (b) show the similarity of the correlation function matrices $C_(x)$ for ideal and noisy digitized circuits of the three-generation Cayley tree-like lattice for regimes of noise, $\epsilon$.}
	\label{SM:Fig:ErrorGate}
\end{figure}

\revisionRefA{
\section{Simulation of gate errors}  \label{Sec:GateError}
In this section we present the gate errors analysis and its impact on our results. To this end, we simulate the digital circuit \textit{with} and \textit{without} gate error, and compare the results expected in each case. To simulate systematic gate error of a given unitary $\hat{U}$, we assume the toy model of a coherent error rotation around an random direction $\vec{d}$ of an small random angle $\epsilon$, such that the imperfect gate $\hat{U}_{\mathrm{imp}}$ reads as
\begin{align}
\hat{U}_{\mathrm{imp}} = \hat{R}_{\vec{d}}(\epsilon)\hat{U}\hat{R}^{\dagger}_{\vec{d}}(\epsilon), ~~ \mathrm{with~} \hat{R}_{\vec{d}}(\epsilon) = e^{i\epsilon \vec{d} \cdot \vec{\sigma}},\nonumber
\end{align}
where $\vec{\sigma} = \sigma_{x}\hat{\i} +  \sigma_{y}\hat{\j} +  \sigma_{z}\hat{\mathrm{k}}$. In order to account the gate error $e_{\mathrm{gate}}$ shown in Table~\ref{table:Device}, we assume that $\epsilon$ as the average error of the gate (in percentage). We exploit the random nature of the vector $\vec{d}\in\Rmath^3$ to introduce the error in our model. If $\hat{U}$ is a single qubit gate, we use the average error for single qubit gates, and for $\hat{U}$ given by the CZ gate we use the average gate error as those ones for two-qubit gates. In our definition $\epsilon$ is an arbitrary parameter to compare different regimes of errors, with $\epsilon = 0$ the ideal digitization protocol. We also consider here the cases where $\epsilon$ = $0.5\%$,$1\%$, and $2\%$. In addition to the coherent gate error, we also simulate dephasing and damping through the standard Kraus operators representation for decohering quantum channels~\cite{Nielsen:Book}, where the decay and dephasing rates are considered identical to all qubits and given by the average of the values given in Table~\ref{table:Device}.

In Fig.~\ref{SM:Fig:ErrorGate} we show three different quantities to qualitatively describe the impact of the errors in our digital algorithm. We consider the similarity of the correlation function of the output state provided by the digitized algorithm after each block, as defined in main text. The two-point correlation function matrix $C$ with elements, $C_{(i,j)} = \langle \hat{\sigma}^{x}_{i}\hat{\sigma}^{x}_{j} \rangle - \langle \hat{\sigma}^{x}_{i} \rangle \langle \hat{\sigma}^{x}_{j} \rangle$. So, we define the correlation matrices $C_{\mathrm{ad}}$, $C_{\mathrm{dig}}^{\mathrm{ideal}}$ and $C_{\mathrm{dig}}^{\mathrm{noisy}}$ associated to the ideal adiabatic solution (analog evolution), digitized optimal evolution and digitized evolution with noise, respectively, we compute $\mathcal{S}(C_{\mathrm{dig}}^{\mathrm{ideal}},C_{\mathrm{ad}})$ and $\mathcal{S}(C_{\mathrm{dig}}^{\mathrm{ideal}},C_{\mathrm{dig}}^{\mathrm{noise}})$. 
The main result from our analysis is shown in Fig.~\ref{SM:Fig:ErrorGate}. We observe that by increasing $\epsilon$ up to $2\%$, the fidelity of the noisy digitization may be drastically affected. But, if we have a gate errors suppressing of $50\%$, we should get enhanced performance of the noisy circuit. In particular, we also show what happens if we can build up a quantum chip satisfying the predictions of minimum gate error for Quantum Advantage in Ref.~\cite{Boixo:18}. In this reference the authors showed that high accuracy computation with digitized quantum processors can be achieved for quantum processors in which two-qubit gate error is around $0.5\%$ and single-qubit error is around $0.05\%$. Our device provides average two-qubit gate error around $1.32\%$ and single-qubit ones around $0.08\%$. It means that by reducing the two-qubit and single-qubit gate error in $65\%$ and $50\%$, respectively, it would be enough to get high fidelity digitization. It is in fact observed in Fig.~\SubFig{SM:Fig:ErrorGate}{b} through the curve denoted by ``Quantum Advantage" limit (QA limit).
}

\begin{figure}[h!]
	\centering
	\includegraphics[width=\linewidth]{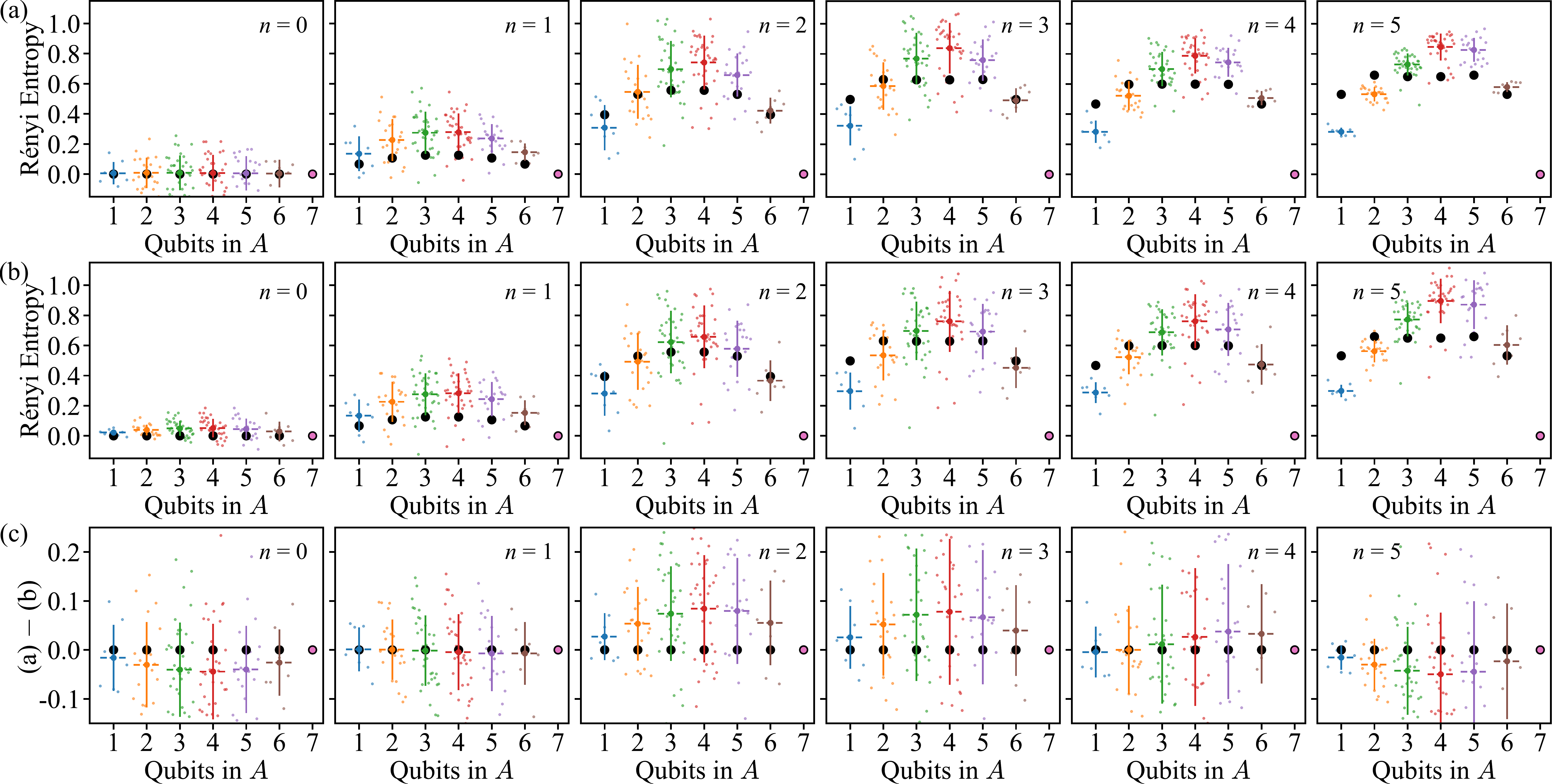}
	\caption{\textbf{Experimental and simulated Rényi entropies.} The Rényi entropy obtained after noise correction and its simulated (ideal) counterpart (black dots) for (a) the ground state and (b) excited states. The difference between (a) and (b) is shown in (c). The parameters for these results are the same as in Fig.~\ref{SM:Fig:BTN-CorrelationFunctios}.}
	\label{SM:Fig:BTN-RenyiNoisy}
\end{figure}

\revisionRefC{
	\section{The Rényi entropy: theory and experimental noise correction}  \label{Sec:GateError2}

In order to show how the error correction considered in the main text help us to get closer to the theoretical (ideal) scenario, we simulated the Rényi entropy for the digital circuit without any noise and we get the new results in Fig.~\ref{SM:Fig:BTN-RenyiNoisy}. In this figure, the same procedure as the experiment is done, but without randomized measurement because we have access to the theoretical full density matrix. 

The black dots are the average of the theoretical Rényi entropy computed for all combinations of the subsystems with $N_{A}$ qubits, as detailed in the main text. From the results in Fig.~\ref{SM:Fig:BTN-RenyiNoisy}, it is possible to verify that our proposed method effectively mitigates the impact of noise-induced errors on entanglement entropy calculations. It is important to mention that the small discrepancy between the experimental and theory data is also explained by the fact that the experimental value are not obtained from the full density matrix, but using a limited amount of random measurements. Even so, it is worth to highlight that such random measurements can characterize entropy with limited computational resources. 

Also, in order to corroborate with the discussion done in the main text, we observe that the quantum correlations profiles are similar through the difference between the data in Fig.~\SubFig{SM:Fig:BTN-RenyiNoisy}{a} and Fig.~\SubFig{SM:Fig:BTN-RenyiNoisy}{b}, where the difference between two data distribution $A$ and $B$ is given as $A-B$. The result is shown in Fig.~\SubFig{SM:Fig:BTN-RenyiNoisy}{c}.
}

\newpage

\bibliography{mybib-URL.bib}
\bibliographystyle{PRA-WithTitle.bst}

